\documentclass[prd,eqsecnum,twocolumn,amsfonts,amssymb]{revtex4}

\usepackage{graphicx}

\usepackage{bm}

\newcommand{\beq}{\begin{equation}}
\newcommand{\eeq}{\end{equation}}
\newcommand{\beqs}{\begin{eqnarray}}
\newcommand{\eeqs}{\end{eqnarray}}

\begin{document}

\title{Infrared Fixed Point Physics in ${\rm SO}(N_c)$ and
 ${\rm Sp}(N_c)$ Gauge Theories}

\author{Thomas A. Ryttov$^a$ and Robert Shrock$^b$}

\affiliation{(a) \ CP$^3$-Origins and Danish Institute for Advanced Study \\
University of Southern Denmark, Campusvej 55, Odense, Denmark}

\affiliation{(b) \ C. N. Yang Institute for Theoretical Physics \\
Stony Brook University, Stony Brook, NY 11794, USA }

\begin{abstract}

  We study properties of asymptotically free vectorial gauge theories with
  gauge groups $G={\rm SO}(N_c)$ and $G={\rm Sp}(N_c)$ and $N_f$ fermions in a
  representation $R$ of $G$, at an infrared (IR) zero of the beta function,
  $\alpha_{IR}$, in the non-Abelian Coulomb phase.  The fundamental, adjoint,
  and rank-2 symmetric and antisymmetric tensor fermion representations are
  considered. We present scheme-independent calculations of the anomalous
  dimensions of (gauge-invariant) fermion bilinear operators 
  $\gamma_{\bar\psi\psi,IR}$ to $O(\Delta_f^4)$ and of the derivative of the
  beta function at $\alpha_{IR}$, denoted $\beta'_{IR}$, to $O(\Delta_f^5)$,
  where $\Delta_f$ is an $N_f$-dependent expansion variable. It is shown that
  all coefficients in the expansion of $\gamma_{\bar\psi\psi,IR}$ that we
  calculate are positive for all representations considered, so that to
  $O(\Delta_f^4)$, $\gamma_{\bar\psi\psi,IR}$ increases monotonically with
  decreasing $N_f$ in the non-Abelian Coulomb phase. Using this property, we
  give a new estimate of the lower end of this phase for some specific
  realizations of these theories.

\end{abstract}

\maketitle


\section{Introduction}
\label{intro_section}

The evolution of an asymptotically free gauge theory from the ultraviolet (UV)
to the infrared is of fundamental importance. The evolution of the running
gauge coupling $g=g(\mu)$, as a function of the Euclidean momentum scale,
$\mu$, is described by the renormalization-group (RG) beta function, $\beta_g =
dg/dt$, or equivalently, $\beta_\alpha = d\alpha/dt$, where $\alpha(\mu) =
g(\mu)^2/(4\pi)$ and $dt=d\ln \mu$ (the argument $\mu$ will often be suppressed
in the notation).  The asymptotic freedom (AF) property means that the gauge
coupling approaches zero in the deep UV, which enables one to perform reliable
perturbative calculations in this regime.  Here we consider a vectorial,
asymptotically free gauge theory (in four spacetime dimensions) with two types
of gauge groups, namely the orthogonal group, $G={\rm SO}(N_c)$, and the
symplectic group (with even $N_c$), $G={\rm Sp}(N_c)$, and $N_f$ copies
(``flavors'') of Dirac fermions transforming according to the respective
(irreducible) representations $R$ of the gauge group, where $R$ is the
fundamental ($F$), adjoint ($A$), or rank-2 symmetric ($S_2$) or antisymmetric
($A_2$) tensor. It may be recalled that for SO($N_c)$, the adjoint and $A_2$
representations are equivalent, while for Sp($N_c)$, the adjoint and $S_2$
representations are equivalent. For technical convenience, we take the fermions
to be massless \cite{fm}. In the case of SO($N_c$), we do not consider 
$N_c=2$, since ${\rm SO}(2) \cong {\rm U}(1)$, and a U(1) gauge theory is 
not asymptotically free (but instead is infrared-free).

If $N_f$ is sufficiently large (but less than the upper limit implied by
asymptotic freedom), then the beta function has an IR zero, at a coupling
denoted $\alpha_{IR}$, that controls the UV to IR evolution \cite{b1,b2}.  
Given that this is the case, as the 
Euclidean scale $\mu$ decreases from the UV to the IR, $\alpha(\mu)$ increases
toward the limiting value $\alpha_{IR}$, and the IR theory is in a chirally
symmetric (deconfined) non-Abelian Coulomb phase (NACP) \cite{sksb}. Here the
value $\alpha = \alpha_{IR}$ is an exact IR fixed point of the renormalization
group, and the corresponding theory in this IR limit is scale-invariant and
generically also conformal invariant \cite{scalecon}. 

The physical properties of the conformal field theory at $\alpha_{IR}$ are of
considerable interest. These properties clearly cannot depend on the scheme
used for the regularization and renormalization of the theory. (We restrict
here to mass-independent schemes.) In usual perturbative calculations,
one computes a given quantity as a series expansion in powers of $\alpha$ to
some finite $n$-loop order.  With this procedure, the result is
scheme-dependent beyond the leading term(s). For example, the beta function is
scheme-dependent at loop order $\ell \ge 3$ and the terms in an anomalous
dimension are scheme-dependent at loop order $\ell \ge 2$ \cite{gross75}. This
applies, in particular, to the evaluation at an IR fixed point. A key fact is
that as $N_f$ (considered to be extended from positive integers to positive
real numbers) approaches the upper limit allowed by the requirement of
asymptotic freedom, denoted $N_u$ (given in Eq. (\ref{nfb1z}) below), it
follows that $\alpha_{IR} \to 0$. Consequently, one can express a physical
quantity evaluated at $\alpha_{IR}$ in a manifestly scheme-independent way as a
series expansion in powers of the variable
\beq
\Delta_f = N_u-N_f \ . 
\label{deltaf}
\eeq
For values of $N_f$ in the non-Abelian Coulomb phase such that $\Delta_f$ is
not too large, one may expect this expansion to yield reasonably accurate
perturbative calculations of physical quantities at $\alpha_{IR}$
\cite{bz}. Some early work on this type of expansion was reported in
\cite{bz,gkgg}. In \cite{gtr}-\cite{dexl} we have presented scheme-independent
calculations of a number of physical quantities at an IR fixed point in an
asymptotically free vectorial gauge theory with a general (simple) gauge group
$G$ and $N_f$ massless fermions in a representation $R$ of $G$, including the
anomalous dimension of a (gauge-invariant) bilinear fermion operator up to
$O(\Delta_f^4)$ and the derivative of the beta function at $\alpha_{IR}$,
$\frac{d\beta}{d\alpha}|_{\alpha=\alpha_{IR}} \equiv \beta'_{IR}$, up to
$O(\Delta_f^5)$. These results for general $G$ and $R$ were evaluated for
$G={\rm SU}(N_c)$ with several fermion representations. Since the global
chiral symmetry is realized exactly in the non-Abelian Coulomb phase, the
bilinear fermion operators can be classified according to their representation
properties under this symmetry, including flavor-singlet and flavor-nonsinglet.
Let $\gamma_{\bar\psi\psi}$ denote the anomalous dimension of the
(gauge-invariant) fermion bilinear, $\bar\psi\psi$ and let
$\gamma_{\bar\psi\psi,IR}$ denote its value at the IR fixed point.  The
scheme-independent expansion of $\gamma_{\bar\psi\psi,IR}$ can be written as
\beq
\gamma_{\bar\psi\psi,IR} = \sum_{j=1}^\infty \kappa_j \, \Delta_f^{\ j}  \ . 
\label{gamma_delta_series}
\eeq
We denote the truncation of right-hand side of Eq. (\ref{gamma_delta_series})
so the upper limit on the sum over $j$ is the maximal power $p$ rather than
$\infty$ as $\gamma_{\bar\psi\psi,IR,\Delta_f^p}$.  The anomalous dimension
$\gamma_{\bar\psi\psi,IR}$ is the same for the flavor-singlet and
flavor-nonsinglet fermion bilinears \cite{gracey_op}, and hence we use the
simple notations $\gamma_{\bar\psi\psi,IR}$ and $\kappa_j$ for both.

The coefficients $\kappa_1$ and $\kappa_2$ are manifestly positive for any $G$
and $R$ \cite{gtr}, and we found that for $G={\rm SU}(N_c)$, $\kappa_3$ and
$\kappa_4$ are also positive for all of the representations $R$ that we
considered \cite{gsi}-\cite{dexl},\cite{nonpos}. This finding implied two
monotonicity results for $G={\rm SU}(N_c)$ and these $R$ and for the range $1
\le p \le 4$ where we had performed these calculations, namely: (i)
$\gamma_{\bar\psi\psi,IR,\Delta_f^p}$ increases monotonically as $N_f$
decreases from $N_u$ in the non-Abelian Coulomb phase; (ii) for a fixed $N_f$
in the NACP, $\gamma_{\bar\psi\psi,IR,\Delta_f^p}$ increases monotonically with
$p$.  We noted that these results in \cite{gtr}-\cite{dexl} motivated the
conjecture that in a (vectorial, asymptotically free) gauge theory with a
general (simple) gauge group $G$ and $N_f$ fermions in a representation $R$ of
$G$, the $\kappa_j$ are positive for all $j$, so that the monotonicity
properties (i) and (ii) would hold for any $p$ in the $\Delta_f$ expansion and
hence also (iii) for fixed $N_f$ in the NACP,
$\gamma_{\bar\psi\psi,IR,\Delta_f^p}$ is a monotonically increasing function of
$p$ for all $p$; (iv) $\gamma_{\bar\psi\psi,IR,\Delta_f^p}$ increases
monotonically as $N_f$ decreases from $N_u$; and (v) the anomalous dimension
$\gamma_{\bar\psi\psi,IR}$ defined by Eq. (\ref{gamma_delta_series}) increases
monotonically with decreasing $N_f$ in the NACP. Clearly, one is motivated to
test this conjecture concerning the positivity of the $\kappa_j$ for other
groups $G$ and fermion representations $R$. Since $\kappa_1$ and $\kappa_2$ are
manifestly positive for any $G$ and $R$, our conjecture on the positivity of
the $\kappa_j$ only needs further testing for the range $j \ge 3$.

In this paper we report our completion of this task for the gauge groups ${\rm
  SO}(N_c)$ and ${\rm Sp}(N_c)$, with fermions transforming
according to the (irreducible) representations $R$ listed above, namely $F$,
$A$, $S_2$, and $A_2$. In the Cartan classification of Lie algebras, $A_n =
{\rm SU}(n+1)$, $B_n = {\rm SO}(2n+1)$, $C_n = {\rm Sp}(2n)$, and $D_n = {\rm
  SO}(2n)$.  For SO($N_c$) with even $N_c$, we restrict to $N_c \ge 6$ since
the algebra $D_n$ is simple if $n \ge 3$, and for Sp($N_c$), we restrict to
even $N_c$, owing to the $D_n = {\rm Sp}(2n)$ correspondence of Lie algebras. 
Henceforth, these restrictions on $N_c$ will be implicit.  We calculate
the coefficients $\kappa_j$ to $O(\Delta_f^4)$ in the $\Delta_f$ series
expansion of the anomalous dimension $\gamma_{\bar\psi\psi,IR}$ of the
(gauge-invariant) fermion bilinear $\bar\psi\psi$.  Again, this is the same for
the flavor-singlet and flavor-nonsinglet bilinears \cite{gracey_op}, so we use
the same notation for both. Stating our results at the outset, we find that (in
addition to the manifestly positive $\kappa_1$ and $\kappa_2$) $\kappa_3$ and
$\kappa_4$ are positive for both the SO($N_c$) and Sp($N_c$) theories and for
all of the representations that we consider.  Some earlier work on the
conformal window in SO($N_c$) and Sp($N_c$) gauge theories, including estimates
of the lower end of this conformal window from perturbative four-loop results
and Schwinger-Dyson methods, was reported in
\cite{sannino_son_spn,sannino_symplectic}. 

We will also use our calculation of $\gamma_{\bar\psi\psi,IR}$ to estimate the
value of $N_f$ that defines the lower end of the non-Abelian Coulomb phase.  We
do this by combining the monotonic behavior that we find for
$\gamma_{\bar\psi\psi,IR,\Delta_f^p}$ for all $p$ that we calculate with an
upper bound on this anomalous dimension from conformal invariance, namely that
$\gamma_{\bar\psi\psi,IR} \le 2$ \cite{gammabound} (discussed further below).
Finally, in addition to our results on $\kappa_j$, we also calculate the
corresponding coefficients $d_j$ in the $\Delta_f$ series expansion of
$\beta'_{IR}$ to $O(\Delta_f^5)$.

Before proceeding, we note that some perspective on these topics can be
obtained from analysis of a vectorial, asymptotically free gauge theory with
${\cal N}=1$ supersymmetry ($ss$) with a gauge group $G$ and $N_f$ pairs of
massless chiral superfields $\Phi$ and $\tilde \Phi$ in the respective
representations $R$ and $\bar R$ of $G$. Here, the upper bound on $N_f$ from
the requirement of asymptotic freedom is $N_{u,ss}=3C_A/(2T_f)$, where $C_A$
and $T_f$ are group invariants (see Appendix \ref{group_invariants}). For this
theory, one can take advantage of a number of exact results
\cite{seiberg,susyreviews}.  These include a determination of the range in
$N_f$ occupied by the non-Abelian Coulomb phase, namely $N_{u,ss}/2 < N_f <
N_{u,ss}$ \cite{nfintegral}, and an exact (scheme-independent) expression for
the anomalous dimension $\gamma_{M,IR}$ of the gauge-invariant bilinear fermion
operator product occurring in the quadratic chiral superfield operator product
$\tilde \Phi \Phi$ at the IR zero of the beta function in the NACP
\cite{susyreviews} (equivalent to $\gamma_{\bar\psi\psi,IR}$ in the
non-supersymmetric theory) namely
\beqs
\gamma_{M,ss}&=&\frac{3C_A}{2T_f N_f}-1
=\frac{N_{u,ss}}{N_f} - 1 \cr\cr
&=& \frac{1}{1-\frac{\Delta_f}{N_{u,ss}}}-1 =
          \sum_{j=1}^\infty \bigg ( \frac{\Delta_f}{N_{u,ss}} \bigg )^j \ . 
\label{gamma_ir_sgt}
\eeqs
As is evident from Eq. (\ref{gamma_ir_sgt}), the
coefficient $\kappa_{j,ss}$ in this supersymmetric gauge theory is 
\beq
\kappa_{j,ss} = \frac{1}{(N_{u,ss})^j} \ , 
\label{kappajss}
\eeq
which is positive-definite for all $j$. To the extent that
one might speculate that this property of the supersymmetric theory could 
carry over to the non-supersymmetric gauge theories considered here, this 
result yields further motivation for our positivity conjecture on the
$\kappa_j$ and the resultant monotonicity properties for the
non-supersymmetric gauge theories that we have given in our earlier work. 
More generally, in \cite{dexss} we calculated exact (scheme-independent)
results for anomalous dimensions of a number of chiral superfield operator
products in a vectorial ${\cal N}=1$ supersymmetric gauge theory \cite{comp}.

This paper is organized as follows. Some relevant background and discussion of
methodology is given in Section \ref{methods_section}. In Sections
\ref{kappaj_section} and \ref{betaprime_section} we present our results for the
$\kappa_j$ and $d_j$ coefficients, respectively.  Our conclusions are given in
Section \ref{conclusions_section} and some relevant group-theoretic inputs are
presented in Appendix \ref{group_invariants}.


\section{Background and Methods} 
\label{methods_section} 

\subsection{Beta Function and Interval $I$ } 

In this section we briefly review some background and methodology relevant for
our calculations. We refer the reader to our previous papers
\cite{gtr}-\cite{dexl} for more details.

The series expansion of $\beta$ in powers of $\alpha$ is
\beq
\beta = -2\alpha \sum_{\ell=1}^\infty b_\ell \,
\Big (\frac{\alpha}{4\pi} \Big )^\ell \ ,
\label{beta}
\eeq
where $b_\ell$ is the $\ell$-loop coefficient.  The truncation of the infinite
series (\ref{beta}) to loop order $\ell=n$ is denoted $\beta_{n\ell}$, and the
physical IR zero of $\beta_{n\ell}$, i.e., the real positive zero closest to
the origin (if it exists) is denoted $\alpha_{IR,n\ell}$. The coefficients
$b_1$ \cite{b1} and $b_2$ \cite{b2} are scheme-independent, while the $b_\ell$
with $\ell \ge 3$ are scheme-dependent \cite{gross75}. 
The higher-loop coefficients $b_\ell$ with
$3 \le \ell \le 5$ have been calculated in \cite{b3}-\cite{b5} (in the 
$\overline{\rm MS}$ scheme \cite{msbar}.)
The conventional expansion of $\gamma_{\bar\psi\psi}$ as a power series in 
the coupling is 
\beq
\gamma_{\bar\psi\psi} = \sum_{\ell=1}^\infty
c_\ell \Big ( \frac{\alpha}{4\pi} \Big )^\ell \ .
\label{gamma}
\eeq
The coefficient $c_1=6C_f$ is scheme-independent, while the $c_\ell$ with $\ell
\ge 3$ are scheme-dependent \cite{gross75}.  The $c_\ell$ were calculated up to
$\ell=4$ in \cite{c4} and to $\ell=5$ in \cite{c5} 
(in the $\overline{\rm MS}$ scheme).

In general, our calculation of the coefficients $\kappa_j$ in the
scheme-independent expansion Eq. (\ref{gamma_delta_series}) requires, as
inputs, the beta function coefficients $b_\ell$ with $1 \le \ell \le j+1$ and
the anomalous dimension coefficients $c_\ell$ with $1 \le \ell \le j$. Because
the $\kappa_j$ are scheme-independent, it does not matter which scheme one uses
to calculate them.  Our calculations used the higher-loop coefficients $b_3$,
$b_4$, and $b_5$ from \cite{b3,b4,b5} and the anomalous dimension coefficients
up to $c_4$ from \cite{c4}.

With a minus sign extracted, as in Eq. (\ref{beta}), the requirement of
asymptotic freedom means that $b_1$ is positive. This condition holds if $N_f$
is less than an upper ($u$) bound, $N_u$, given by the value where $b_1$ is
zero
\beq
N_u = \frac{11C_A}{4T_f} \ .
\label{nfb1z}
\eeq
Hence, the asymptotic freedom condition yields the upper bound $N_f <
N_u$. With the overall minus sign extracted in Eq. (\ref{beta}), the one-loop
coefficient $b_1$ is positive if $N_f < N_u$. 

In the asymptotically free regime, $b_2$ is negative if $N_f$ lies in the
interval $I$
\beq
I: \quad N_{\ell} < N_f < N_u \ ,
\label{nfinterval}
\eeq
where the value of $N_f$ at the lower end is \cite{nfintegral}
\beq
N_{\ell} = \frac{17C_A^2}{2T_f(5C_A+3C_f)} \ .
\label{nfb2z}
\eeq
For $N_f \in I$, the two-loop beta function has an IR zero, which occurs at the
value $\alpha_{IR,2\ell}=-4\pi b_1/b_2$.  As $N_f$ approaches $N_u$ from below,
the IR zero of the beta function goes to zero.  As $N_f$ decreases below $N_u$,
the value of this IR zero increases, motivating its calculation to higher
order.  This has been done up to four-loop order in \cite{bvh}-\cite{bc} and up
to five-loop order in \cite{flir}.  The scheme-dependence has been studied in
\cite{sch}-\cite{gracey2015}. For a given $G$ and $R$, the value of $N_f$ below
which the gauge interaction spontaneously breaks chiral symmetry is denoted
$N_{f,cr}$. (Note that $N_{f,cr}$ does not, in general, coincide with
$N_\ell$.)


\subsection{Interval $I$ for Specific $R$}

We proceed to list explicit expressions for the upper and lower ends of the
interval $I$ where the two-loop beta function has an IR zero, and associated
quantities for the representations of SO($N_c$) and Sp($N_c$) under
consideration here. It will be convenient to list these together, with the
understanding that the upper and lower signs refer to SO($N_c$) and
Sp($N_c$), respectively.


\subsubsection{$R=F$} 

For the fundamental representation, $R=F$, Eqs. (\ref{nfb1z}) and 
(\ref{nfb2z}) yield
\beq
N_{u,F} = \frac{11(N_c \mp 2)}{4}
\label{nfb1z_fund}
\eeq
and
\beq
N_{\ell,F} = \frac{17(N_c \mp 2)^2}{13N_c \mp 23} \ . 
\label{nfb2z_fund}
\eeq
Thus, the intervals $I$ in which the two-loop beta function has an IR zero for
this case $R=F$ for these two respective theories are 
\beqs
R=F: \quad 
I: \quad \frac{17(N_c \mp 2)^2}{13N_c \mp 23} < N_f < \frac{11(N_c \mp 2)}{4}
\ . \cr\cr
&&
\label{interval_fund}
\eeqs

The maximum values of $\Delta_{f,F} = N_{u,F}-N_f$ for $N_f \in I$ for these
theories are 
\beq
\Delta_{f,max,F} = \frac{3(N_c \mp 2)(25N_c \mp 39)}{4(13N_c \mp 23)} \ . 
\label{deltaf_max_fund}
\eeq
%


\subsubsection{LNN Limit} 

For this $R=F$ case, it is of interest to consider the limit 
\beqs
& & LNN: \quad N_c \to \infty \ , \quad N_f \to \infty \cr\cr
& & {\rm with} \ r \equiv \frac{N_f}{N_c} \ {\rm fixed \ and \ finite}  \cr\cr
& & {\rm and} \ \ \xi(\mu) \equiv \alpha(\mu) N_c \ {\rm is \ a \
finite \ function \ of} \ \mu \ .
\cr\cr
& &
\label{lnn}
\eeqs
As in our earlier work, we use the symbol $\lim_{LNN}$ for this limit (also
called the 't Hooft-Veneziano limit), where ``LNN'' stands for ``large $N_c$
and $N_f$'' with the constraints in Eq. (\ref{lnn}) imposed.  One of the useful
features of the LNN limit is that, for a general gauge group $G$ and a given
fermion representation $R$ of $G$, one can make $\alpha_{IR}$ arbitrarily small
by analytically continuing $N_f$ from the nonnegative integers to the real
numbers and letting $N_f \to N_u$.

We define 
\beq
r_u = \lim_{LNN} \frac{N_u}{N_c} \ ,
\label{rb1zdef}
\eeq
and
\beq
r_\ell = \lim_{LNN} \frac{N_\ell}{N_c} \ ,
\label{rb2zdef}
\eeq
The critical value of $r$ such that for $r > r_{cr}$, the
LNN theory is in the non-Abelian Coulomb phase and hence is inferred to be 
IR-conformal is denoted $r_{cr}$ and is defined as
\beq
r_{cr} = \lim_{LNN} \frac{N_{f,cr}}{N_c} \ .
\label{rcr}
\eeq
We define the scaled scheme-independent expansion parameter in this LNN limit
as 
\beq
\Delta_r \equiv \frac{\Delta_f}{N_c} = r_u-r  \ . 
\label{deltar}
\eeq
In the LNN limit, the coefficient $\kappa_{j,F}$ has the asymptotic behavior
$\kappa_{j,F} \propto 1/N_c^j + O(1/N_c^{j+1})$.  Consequently, the
quantities that are finite in this limit are the rescaled coefficients 
\beq
\hat\kappa_{j,F} \equiv \lim_{LNN} N_c^j \kappa_{j,F} \ . 
\label{kappajhat}
\eeq
The anomalous dimension
$\gamma_{\bar\psi\psi,IR}$ is finite in this limit and is given by
\beq
R=F: \quad \lim_{LNN} \gamma_{\bar\psi\psi,IR} 
= \sum_{j=1}^\infty \kappa_{j,F} \Delta_f^j
= \sum_{j=1}^\infty \hat \kappa_{j,F} \Delta_r^j \ .
\label{gamma_ir_lnn}
\eeq

In the LNN limit, for both the SO($N_c$) and Sp($N_c$) theories, 
\beq
LNN: \quad r_u = \frac{11}{4}, \quad r_\ell = \frac{17}{13},
\label{rvalues_fund}
\eeq
and the resultant interval $I_r$, $r_\ell < r < r_u$, is
\beq
LNN: \quad 
\frac{17}{13} < r < \frac{11}{4}, \quad i.e., \quad 1.3077 < r < 2.750
\label{intervalr_fund}
\eeq
The maximum value, $\Delta_{r,max} = r_u-r$ for $r \in I_r$ is
\beq
LNN: \quad \Delta_{r,max} = r_u-r_\ell = \frac{75}{52} = 1.4423
\label{deltar_max_lnn}
\eeq
%


\subsubsection{$R=A$}

For fermions in the adjoint representation, $R=A$, of both the SO($N_c$) and 
Sp($N_c$) theories Eqs. (\ref{nfb1z}) and (\ref{nfb2z}) take the form
\beq
N_{u,A} = \frac{11}{4} 
\label{nfb1z_adj}
\eeq
and
\beq
N_{\ell,A} = \frac{17}{16} \ , 
\label{nfb2z_adj}
\eeq
so that the interval $I$ for both of these theories is
\beq
R=A \ \Rightarrow \ I: \quad \frac{17}{16} < N_f < \frac{11}{4} \ , 
\label{interval_adj}
\eeq
i.e., $1.0625 < N_f < 2.750$.  This interval includes only one physical,
integral value of $N_f$, namely $N_f=2$. With a formal generalization of $N_f$
from positive integral to real values, the maximal value of $\Delta_{f,A}$ 
for $N_f \in I$ is 
\beq
\Delta_{f,max,A} = \frac{27}{16} = 1.6875
\label{deltaf_max_adj}
\eeq
As noted above, the $A$ and $A_2$ representations are equivalent in SO($N_c$),
and the $A$ and $S_2$ representations are equivalent in Sp($N_c$).

For this $R=A$ case, it is also be of interest to
consider the original 't Hooft limit, denoted here as the LN (``large $N_c$'')
limit, namely 
\beqs
&& LN: \quad N_c \to \infty \cr\cr
&& {\rm with} \ \xi(\mu) \equiv \alpha(\mu) N_c \ {\rm a \
finite \ function \ of} \ \mu \cr\cr
&&
\label{ln}
\eeqs
and $N_f$ fixed and finite.


\subsubsection{$R=S_2$ for SO($N_c$) and $R=A_2$ for Sp($N_c$) }

For the symmetric rank-2 tensor representation of SO($N_c$), $S_2$,
Eqs. (\ref{nfb1z}) and (\ref{nfb2z}) reduce to
\beq
N_{u,S_2,{\rm SO}(N_c)} = \frac{11(N_c-2)}{4(N_c+2)} 
\label{nfb1z_sym_so}
\eeq
and
\beq
N_{\ell,S_2,{\rm SO}(N_c)} = \frac{17(N_c-2)^2}{4(N_c+2)(4N_c-5)} . 
\label{nfb2z_sym_so}
\eeq
Since $N_{u,S_2,{\rm SO}(N_c)} < 1$ if $N_c < 30/7 = 4.286$, it follows that if
$N_c=3$ or $N_c=4$, then the asymptotic freedom condition forbids an SO($N_c$)
theory from having any fermion in the $S_2$ representation.  As $N_c$ increases
through the value 30/7, the upper bound on the number $N_f$ from asymptotic
freedom, $N_{u,S_2,{\rm SO}(N_c)}$, increases through unity, and as $N_c$
increases through the value $38/3 = 12.667$, $N_{u,S_2,{\rm SO}(N_c)}$
increases through the value 2.  As $N_c \to \infty$, $N_{u,S_2,{\rm SO}(N_c)}$
approaches the limit 11/4 = 2.75 from below.  Hence, for physical
integral values of $N_c$, in the range $5 \le N_c \le 12$, an asymptotically
free SO($N_c$) theory may have at most $N_f=1$ fermion in the $S_2$
representation, and for $N_c \ge 13$, this theory may have at most $N_f=2$
fermions in the $S_2$ representation. The lower boundary of the interval $I$,
$N_{\ell,S_2,{\rm SO}(N_c)}$, is a monotonically increasing function of $N_c$
which increases through unity as $N_c$ increases through the value
$N_c=2(20+\sqrt{373} \ ) = 78.626$ and approches the limit 17/16=1.0625 as $N_c
\to \infty$. Hence, for integral $N_c \ge 79$, the interval $I$ for SO($N_c$)
only contains the single value $N_f=2$.

The maximum value of $\Delta_{f,S_2} = N_{u,S_2}-N_{\ell,S_2}$ for SO($N_c$)
and $N_f \in I$ is 
\beq
\Delta_{f,max,S_2,{\rm SO}(N_c)}=\frac{3(N_c-2)(9N_c-7)}{4(N_c+2)(4N_c-5)} \ .
\label{deltaf_max_sym_so}
\eeq
%


\subsubsection{$R=A_2$ for Sp($N_c$) }

We next consider the antisymmetric rank-2 representation of Sp($N_c$),
$A_2$. This is a singlet for $N_c=2$, so in the present discussion we restrict
to (even) $N_c \ge 4$. We have 
\beq
N_{u,A_2,{\rm Sp}(N_c)} = \frac{11(N_c+2)}{4(N_c-2)} 
\label{nfb1z_asym_sp}
\eeq
and
\beq
N_{\ell,A_2,{\rm Sp}(N_c)} = \frac{17(N_c+2)^2}{4(N_c-2)(4N_c+5)} . 
\label{nfb2z_asym_sp}
\eeq
Both $N_{u,A_2,{\rm Sp}(N_c)}$ and $N_{\ell,A_2,{\rm Sp}(N_c)}$ decrease
monotonically in the relevant range of (even) $N_c \ge 4$ for this theory,
approaching the respective limits 11/4 and 17/16 as $N_c \to \infty$. 
The maximum value of $\Delta_{f,A_2} = N_{u,A_2}-N_{\ell,A_2}$ for 
Sp($N_c$) and $N_f \in I$ is 
\beq
\Delta_{f,max,A_2,{\rm Sp}(N_c)}=\frac{3(N_c+2)(9N_c+7)}{4(N_c-2)(4N_c+5)} \ .
\label{deltaf_max_asym_sp}
\eeq
These results for $R=A_2$ in Sp($N_c$) are simply related by sign
reversals of various terms to the results for $R=S_2$ in SO($N_c$). 


\subsection{Conformality Upper Bound on Anomalous Dimension}

We denote the full scaling dimension of a (gauge-invariant) quantity ${\cal O}$
as $D_{\cal O}$ and its free-field value as $D_{{\cal O},free}$.  The anomalous
dimension of this operator, denoted $\gamma_{\cal O}$, is defined via the
equation \cite{gammaconvention} 
\beq
D_{\cal O} = D_{{\cal O},free} - \gamma_{\cal O} \ .
\label{anomdim}
\eeq
Operators of
particular interest include fermion bilinears of the form $\bar\psi\psi =
\bar\psi_R \psi_L + \bar\psi_L \psi_R$, where it is understood that gauge
indices are contracted in such a way as to yield a gauge-singlet. As discussed
above, the anomalous dimension at the IR fixed point,
$\gamma_{\bar\psi\psi,IR}$, is scheme-independent and is the same for
flavor-singlet and flavor-nonsinglet operators \cite{gracey_op}, and hence we
suppress the flavor indices in the notation.  

There is a lower bound on the full dimension of a Lorentz-scalar operator
${\cal O}$ (other than the identity) in a conformally invariant theory, which
is $D_{\cal O} \ge 1$ \cite{gammabound}. With the definition (\ref{anomdim}),
this is equivalent to the upper bound on the anomalous dimension of ${\cal
  O}$. For the non-supersymmetric theories considered in this paper, this is
the upper bound
\beq
\gamma_{\bar\psi\psi,IR} \le 2 \ . 
\label{gamma_upperbound}
\eeq
For the gauge-invariant fermion bilinear occuring in the quadratic superfield 
operator product in a supersymmetric gauge theory, the analogous upper bound is
1 rather than 2, since $\psi$ occurs in conjunction with the Grassmann 
$\theta$ with dimension $-1/2$ in the chiral superfield (see \cite{dex} for a
more detailed discussion). 

As is evident from Eq. (\ref{gamma_ir_sgt}), the analogue of
$\gamma_{\bar\psi\psi,IR}$ in the supersymmetric theory, namely
$\gamma_{M,IR}$, increases monotonically with decreasing $N_f$ in the
non-Abelian Coulomb phase. Furthermore, it saturates its unitarity upper bound
$\gamma_{M,IR} \le 1$ from conformal invariance at the lower end of the NACP.
At present, one does not know if $\gamma_{\bar\psi\psi,IR}$ in (vectorial,
asymptotically free) non-supersymmetric gauge theories saturates its upper
bound of 2 as $N_f$ decreases to $N_{f,cr}$ in the conformal, non-Abelian
Coulomb phase. Assuming that these monotonicity and saturation properties also
hold for $\gamma_{\bar\psi\psi,IR}$ in the NACP of a (vectorial, asymptotically
free) non-supersymmetric gauge theory, if one had an exact expression for
$\gamma_{\bar\psi\psi,IR}$, then, for a given $G$ and $R$, one could derive the
value of $N_f$ at the lower end of the NACP by setting
$\gamma_{\bar\psi\psi,IR}=2$ and solving for $N_f$ \cite{sd}. In practice, one
can only obtain an estimate of $N_{f,cr}$ in this manner, since one does not
have an exact expression for $\gamma_{\bar\psi\psi,IR}$. One way that this can
be done is via conventional $n$-loop calculations of the zero of the beta
function at $\alpha_{IR,n\ell}$ and the value of $\gamma_{\bar\psi\psi,IR}$ at
this zero, denoted $\gamma_{\bar\psi\psi,IR,n\ell}$, which was done up to the
four-loop level in \cite{bvh,ps} and up to the five-loop level in
\cite{flir}. An arguably better approach is to work with the expansion, in
powers of $\Delta_f$ \cite{bz}, of $\gamma_{\bar\psi\psi,IR}$, since this is
scheme-independent.  We have done this in \cite{gtr,gsi,dex}, and up to order
$O(\Delta_f^4)$ in \cite{dexs,dexl} (using the five-loop beta function, as
noted above). In order to apply this method to estimate $N_{f,cr}$, it is
necessary that all of the coefficients $\kappa_j$ are used for the estimate
must be positive, so that the resultant $\gamma_{\bar\psi\psi,IR,\Delta_f^p}$
monotonically increases with decreasing $N_f$ in the NACP, and this requirement
was satisfied for $G={\rm SU}(N_c)$ and all of the fermion representations $R$
that we used.  As discussed in detail in \cite{gsi,dex,dexs,dexl}, our
estimates of $N_{f,cr}$ from this work are in general agreement, to within the
uncertainties, with estimates from lattice simulations (bearing in mind that,
for the various SU($N_c$) groups and fermion representations $R$, not all
lattice groups agree on the resultant estimate of $N_{f,cr}$).


\subsection{ $\beta'_{IR}$ } 

Another scheme-independent quantity of interest is the derivative of the beta
function at the IR fixed point, $\beta'_{IR}$. This is equivalent to the
anomalous dimension of ${\rm Tr}(F_{\mu\nu}F^{\mu\nu})$ at the IR fixed point,
where $F^a_{\mu\nu}$ is the gluonic field strength tensor \cite{traceanomaly}.
The derivative $\beta'_{IR}$ has the scheme-independent expansion
\beq
\beta'_{IR} = \sum_{j=2}^\infty d_j\Delta_f^{\ j} \ . 
\label{betaprime_delta_series}
\eeq
As indicated, $\beta'_{IR}$ has no term linear in $\Delta_f$.  In general, the
calculation of the scheme-independent coefficient $d_j$ requires, as inputs,
the $b_\ell$ for $1 \le \ell \le j$. Our calculations of $d_j$ for $2 \le j \le
4$ in \cite{dex} used the higher-order coefficients $b_3$ from \cite{b3} and
$b_4$ from \cite{b4}, and our calculations of $d_5$ in \cite{dexs,dexl} used
$b_5$ from \cite{b5su3,b5}. A detailed analysis of the region of convergence
of the series expansions (\ref{gamma_delta_series}) and
(\ref{betaprime_delta_series}) in powers of $\Delta_f$ was given in
\cite{dex,dexs,dexl}, and we refer the reader to these references for a
discussion of this analysis.


\section{Calculation of Coefficients $\kappa_{j,R}$ for SO($N_c$) and 
S\lowercase{p}($N_c$) }
\label{kappaj_section} 

We calculated general expressions for the $\kappa_j$ for a group $G$ and
fermions in a representation $R$ for $1 \le j \le 3$ in \cite{gtr,dex} and for
$j=4$ in \cite{dexs,dexl}. The coefficients $\kappa_1$ and $\kappa_2$ are
manifestly positive, as is evident from their expressions, 
\beq
\kappa_1 = \frac{8C_fT_f}{C_A(7C_A+11C_f)} \ ,
\label{kappa1}
\eeq
\beq
\kappa_2 = \frac{4C_fT_f^2(5C_A+88C_f)(7C_A+4C_f)}{3C_A^2(7C_A+11C_f)^3} \ ,
\label{kappa2}
\eeq
and we found that $\kappa_3$ and $\kappa_4$ were also positive for $G={\rm
  SU}(N_c)$ and all of the fermion representations $R$ that we considered,
which included the fundamental, adjoint, and rank-2 symmetric and antisymmetric
tensor representations.  As noted above, one of the main goals of the present
work is to determine if this positivity also holds for SO($N_c$) and Sp($N_c$)
theories as well as our established result for SU($N_c$) theories. 


\subsection{$R=F$} 

Because the various group invariants for SO($N_c$) and Sp($N_c$) are simply
related to each other, it is convenient to present our results for these two
theories together.  For fermions in the fundamental representation, our general
formulas reduce to the following explicit expressions, where the upper and
lower signs refer to $G={\rm SO}(N_c)$ and $G={\rm Sp}(N_c)$, respectively:
\beq
\kappa_{1,F} = \frac{2^3(N_c \mp 1)}{(N_c \mp 2)(25N_c \mp 39)} \ , 
\label{kappa1_fund}
\eeq
\bigskip
\beq
\kappa_{2,F} = \frac{2^4(N_c \mp 1)(9N_c \mp 16)(49N_c \mp 54)}
{3(N_c \mp 2)^2(25N_c \mp 39)^3} \ ,
\label{kappa2_fund}
\eeq
\begin{widetext}
\beqs
\kappa_{3,F}&=&
\frac{2^6(N_c \mp 1)}{3^3(N_c \mp 2)^3(25N_c \mp 39)^5}\bigg [ 
\Big ( 274243N_c^4 \mp 1638318N_c^3+3586884N_c^2 \mp 3298968N_c+1018710 \Big )
 \cr\cr
& \pm &2^7 \cdot 33(N_c \mp 3)(3N_c \pm 2)(25N_c \mp 39)\zeta_3 \ \bigg ] \ , 
\label{kappa3_fund}
\eeqs
and
\beqs
\kappa_{4,F} &=& \frac{2^6(N_c \mp 1)}{3^4(N_c \mp 2)^4(25N_c \mp 39)^7} \, 
\Bigg [ \bigg (263345440N_c^6 \mp 2325643530N_c^5+8506782306N_c^4 \cr\cr
&&\mp 16264883388N_c^3+16883765721N_c^2 \mp 8888128812N_c+1834476660\bigg ) 
\cr\cr
&&+2^6(25N_c \mp 39)\bigg ( 26400N_c^5 \pm 235846N_c^4-1427001N_c^3 \pm 
1629821N_c^2 -404418N_c \pm 720594 \bigg )\zeta_3 \cr\cr
&&+2^8 \cdot 275(N_c \mp 2)(N_c \mp 3)(25N_c \mp 39)^2
(3N_c^2 \mp 23N_c-16)\zeta_5 \ \Bigg ] \ ,
\label{kappa4_fund}
\eeqs
\end{widetext} 
where $\zeta_s = \sum_{n=1}^\infty n^{-s}$ is the Riemann zeta function, with
$\zeta_3=1.202057$ and $\zeta_5=1.036928$ (given to the indicated
floating-point accuracy). In addition to $\kappa_1$ and $\kappa_2$, which are
manifestly positive for any (simple) gauge group $G$ and fermion representation
$R$, we find, by numerical evaluation, that $\kappa_{3,F}$ and $\kappa_{4,F}$
are positive for the relevant ranges of $N_c$ in both of these theories. 

\begin{figure}
  \begin{center}
    \includegraphics[height=6cm]{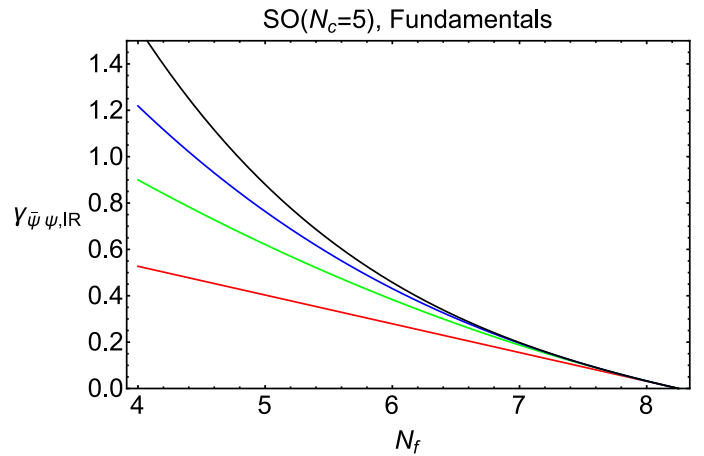}
  \end{center}
  \caption{Plot of $\gamma_{\bar\psi\psi,IR,F,\Delta_f^p}$ (labelled as
    $\gamma_{\bar\psi\psi,IR}$ on the vertical axis) for an SO(5) gauge theory
    with fermions in the fundamental representation $R=F$, with $1 \le p \le
    4$, as a function of $N_f \in I$. From bottom to top, the curves (with
    colors online) refer to $\gamma_{\bar\psi\psi,IR,F,\Delta_f}$ (red),
    $\gamma_{\bar\psi\psi,IR,F,\Delta_f^2}$ (green),
    $\gamma_{\bar\psi\psi,IR,F,\Delta_f^3}$ (blue), and
    $\gamma_{\bar\psi\psi,IR,F,\Delta_f^4}$ (black).}
\label{gamma_SO5_fund_plot}
\end{figure}

As an explicit example of our scheme-independent calculations of
$\gamma_{\bar\psi\psi,IR}$ to $O(\Delta_f^p)$ with $1 \le p \le 4$ for an
SO($N_c$) group, let us consider an SO(5) gauge group with fermions in the
fundamental representation.  For this theory, the general formulas
Eqs. (\ref{nfb1z}) and (\ref{nfb2z}) give $N_{u,F}=33/4 = 8.25$ and
$N_\ell=51/14=3.643$ \cite{nfintegral}, so that, with $N_f$ generalized to real
numbers, the interval $I$ is $3.643 < N_f < 8.125$ of which the physical,
integral values of $N_f$ are given by the interval $4 \le N_f \le 8$.  In
Fig. \ref{gamma_SO5_fund_plot} we present a plot of our $O(\Delta_f^p)$
scheme-independent calculations of $\gamma_{\bar\psi\psi,IR}$, viz.,
$\gamma_{\bar\psi\psi,IR,\Delta_f^p}$, with $1 \le p \le 4$. (The
representations $R=F$ is indicated explicitly in the notation for the figure,
as $\gamma_{\bar\psi\psi,IR,F,\Delta_f^p}$). Combining these
results with our positivity conjecture for higher $p$ and our saturation
assumption and the conformality upper bound (\ref{gamma_upperbound}) yields an
estimate of $N_{f,cr}$ for this SO(5) theory, namely $N_{f,cr} \sim 4$.  This
procedure entails an estimate of an extrapolation of our results for
$\gamma_{\bar\psi\psi,IR,F,\Delta_f^p}$, with $1 \le p \le 4$ to $p=\infty$,
yielding the exact $\gamma_{\bar\psi\psi,F,IR}$ defined by the infinite series
(\ref{gamma_delta_series}).  We remark that this estimated value, $N_{f,cr}
\sim 4$, is close to (and is the integer nearest to) the lower end of the
interval $I$ at $N_f=3.643$. To our knowledge, there has not yet been a
reported lattice measurement of $\gamma_{\bar\psi\psi,F,IR}$ in the non-Abelian
Coulomb phase for this theory, with which our estimate of 
$\gamma_{\bar\psi\psi,IR}$ could be compared. 

\begin{figure}
  \begin{center}
    \includegraphics[height=6cm]{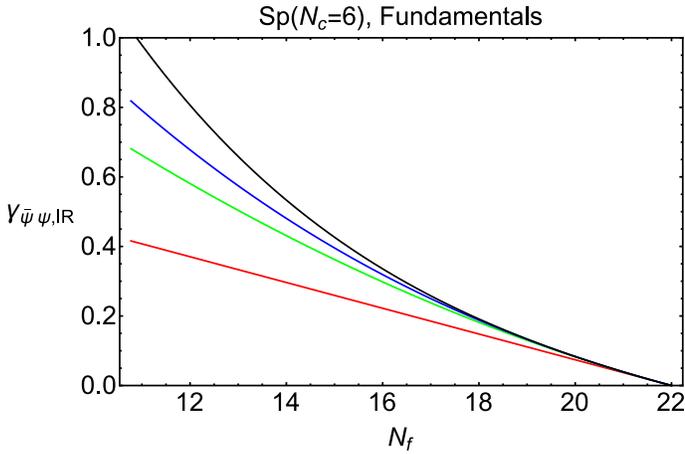}
  \end{center}
  \caption{Plot of $\gamma_{\bar\psi\psi,IR,F,\Delta_f^p}$ (labelled as
    $\gamma_{\bar\psi\psi,IR}$ on the vertical axis) for an Sp(6) gauge theory
    with fermions in the fundamental representation, $R=F$, with $1 \le p \le
    4$, as a function of $N_f \in I$. From bottom to top, the curves (with
    colors online) refer to $\gamma_{\bar\psi\psi,IR,F,\Delta_f}$ (red),
    $\gamma_{\bar\psi\psi,IR,F,\Delta_f^2}$ (green),
    $\gamma_{\bar\psi\psi,IR,F,\Delta_f^3}$ (blue), and
    $\gamma_{\bar\psi\psi,IR,F,\Delta_f^4}$ (black).}
\label{gamma_Sp6_fund_plot}
\end{figure}

Similarly, as an explicit example of our calculations of
$\gamma_{\bar\psi\psi,IR}$ to $O(\Delta_f^p)$ with $1 \le p \le 4$ for an
Sp($N_c$) group, we will consider an Sp(6) gauge group, again with fermions in
the fundamental representation. We choose this example rather than Sp(4)
because of the isomorphism ${\rm SO}(5) \cong {\rm Sp}(4)$ (see Appendix
\ref{group_invariants}.) From Eqs. (\ref{nfb1z}) and (\ref{nfb2z}) we obtain
the values $N_{u,F}=22$ and $N_\ell=1088/101 = 10.772$ \cite{nfintegral}, so
that, with $N_f$ generalized to real numbers, the interval $I$ is $10.772 < N_f
< 22$ of which the physical, integral values of $N_f$ are given by the interval
$11 \le N_f \le 21$.  In Fig. \ref{gamma_Sp6_fund_plot} we present a plot of
our $O(\Delta_f^p)$ scheme-independent calculations of
$\gamma_{\bar\psi\psi,F,IR}$, viz., $\gamma_{\bar\psi\psi,IR,F,\Delta_f^p}$,
with $1 \le p \le 4$. Applying our monotonicity conjecture and estimation
methods in the same way as with the SO(5) example above, we are led to the
inference that $N_{f,cr}$ is somewhat below the lower end of the interval
$I$. As was the case with SO(5), we are not aware of any lattice study of this
theory with which we could compare these inferences.

It is straightforward to use our calculations for $\kappa_j$ in
Eqs. (\ref{kappa1_fund})-(\ref{kappa4_fund}) to compute
$\gamma_{\bar\psi\psi,IR,F,\Delta_f^p}$ with $1 \le p \le 4$ for SO($N_c$) and
Sp($N_c$) theories with $R=F$ and other values of $N_c$, and to make estimates
of the lower end of the NACP for these other $N_c$, but the examples given
above should suffice to illustrate the method.

We next mention some checks on our general calculation of the $\kappa_j$
coefficients for SO($N_c$) and Sp($N_c$) with $R=F$. One has the isomorphism 
${\rm SO}(3) \cong {\rm SU}(2)$, and, as part of this, the 
fundamental representation of SO(3) is equivalent to the adjoint
representation of SU(2).  Hence, 
\beq
\kappa_{j,F,{\rm SO}(3)} = \kappa_{j,A,{\rm SU}(2)} \quad \forall \ j \ , 
\label{kappaj_su2adj_so3fund}
\eeq
where we have indicated the gauge group $G$ and the fermion representation $R$
as subscripts. Using our previous calculations of $\kappa_j$ for the SU($N_c$)
gauge theory with fermions in the adjoint representation, we have verified 
that our present calculation of $\kappa_j$ satisfies this check. Explicitly,
with the different gauge groups indicated explicitly, we have
\beq
\kappa_{1,F,{\rm SO}(3)} = \kappa_{1,A,{\rm SU}(2)} =  \frac{2^2}{3^2} =
0.444444 \ , 
\label{kappa1_so3_fund}
\eeq
\beq
\kappa_{2,F,{\rm SO}(3)} = \kappa_{2,A,{\rm SU}(2)} = 
\frac{341}{2 \cdot 3^6} = 0.233882 \ , 
\label{kappa2_so3_fund}
\eeq
\beq
\kappa_{3,F,{\rm SO}(3)} = \kappa_{3,A,{\rm SU}(2)} = 
\frac{51217}{2^3 \cdot 3^{10}} = 0.108421 \ , 
\label{kappa3_so3_fund}
\eeq
and
\beq
\kappa_{4,F,{\rm SO}(3)} = \kappa_{4,A,{\rm SU}(2)} = 
\frac{47764753}{2^7 \cdot 3^{14}} + \frac{9592}{3^{11}}\zeta_3 = 0.143107 \ .
\label{kappa4_so3_fund}
\eeq

From the explicit expressions above, we calculate the following values of the 
$\hat\kappa_{j,F}$, which are the same in the LNN limits of the SO($N_c$) and
Sp($N_c$) theories (with the
numerical values given to the indicated precision):
\beq
\hat\kappa_{1,F} = \frac{2^3}{5^2} = 0.320000 \ , 
\label{kappa1hat}
\eeq
\beq
\hat\kappa_{2,F} = \frac{2^4 \cdot 147}{5^6} = 0.150528 \ , 
\label{kappa2hat}
\eeq
\beq
\hat\kappa_{3,F} = \frac{2^6 \cdot 274243}{3^3 \cdot 5^{10}} = 0.0665659 \ , 
\label{kappa3hat}
\eeq
and
\beqs
\hat\kappa_{4,F} &=& \frac{2^{11} \cdot 1645909}{3^4 \cdot 5^{13}} 
+ \frac{2^{17} \cdot 11}{3^3 \cdot 5^{10}}\, \zeta_3 + 
\frac{2^{14} \cdot 11}{3^3 \cdot 5^8}\, \zeta_5 \cr\cr
&=& 0.0583830 \ . 
\label{kappa4hat}
\eeqs
Here we have indicated the simple factorizations of the denominators. In
general, the numerators do not have simple factorizations, although they often
contain various powers of 2, as indicated. We shall generally use
this factorization format throughout the paper.


\subsection{ $R=A$ } 

For $R=A$, we find the following coefficients, where again the upper and lower 
signs refer to SO($N_c$) and Sp($N_c$).  The floating-point values are 
quoted to the indicated numerical precision: 
\beq
\kappa_{1,A} = \Big ( \frac{2}{3} \Big )^2 = 0.444444 \ , 
\label{kappa1_adj}
\eeq
\beq
\kappa_{2,A} = \frac{341}{2 \cdot 3^6} = 0.233882 \ , 
\label{kappa2_adj}
\eeq
\beq
\kappa_{3,A} = \frac{61873N_c^3 \mp 360582N_c^2+593292N_c \mp 153992}
{2^3 \cdot 3^{10}(N_c \mp 2)^3} \ , 
\label{kappa3_adj}
\eeq
and
\begin{widetext}
\beqs
\kappa_{4,A} &=& \frac{1}{2^7 \cdot 3^{14} (N_c \mp 2)^3} \, 
\bigg [ \Big ( 53389393N_c^3 \mp 314711718N_c^2+561927756N_c
\mp 247126664 \Big ) \cr\cr
&&+\Big (3815424N_c^3 \mp 52227072N_c^2+456468480N_c \mp 
969228288 \Big )\zeta_3 \ \bigg ] \ . 
\label{kappa4_adj}
\eeqs
\end{widetext} 

\begin{figure}
  \begin{center}
    \includegraphics[height=6cm]{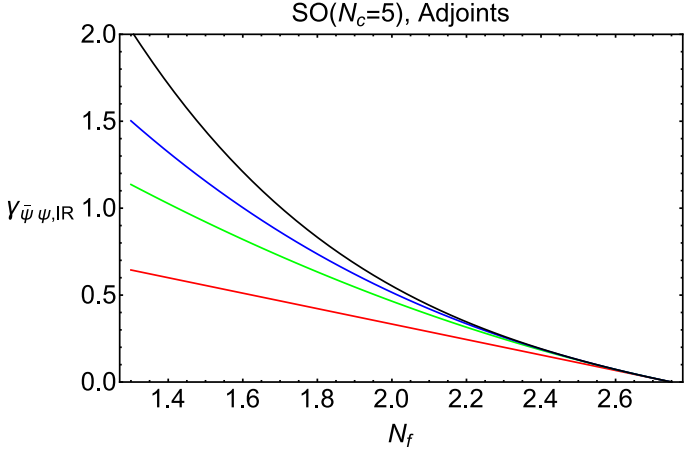}
  \end{center}
  \caption{Plot of $\gamma_{\bar\psi\psi,IR,A,\Delta_f^p}$ (labelled as
    $\gamma_{\bar\psi\psi,IR}$ on the vertical axis) for an SO(5) gauge theory
    with fermions in the adjoint representation $R=A$, with $1 \le p \le 4$, as
    a function of $N_f \in I$. From bottom to top, the curves (with colors
    online) refer to $\gamma_{\bar\psi\psi,IR,A,\Delta_f}$ (red),
    $\gamma_{\bar\psi\psi,IR,A,\Delta_f^2}$ (green),
    $\gamma_{\bar\psi\psi,IR,A,\Delta_f^3}$ (blue), and
    $\gamma_{\bar\psi\psi,IR,A,\Delta_f^4}$ (black).}
\label{gamma_SO5_adj_plot}
\end{figure}

\begin{figure}
  \begin{center}
    \includegraphics[height=6cm]{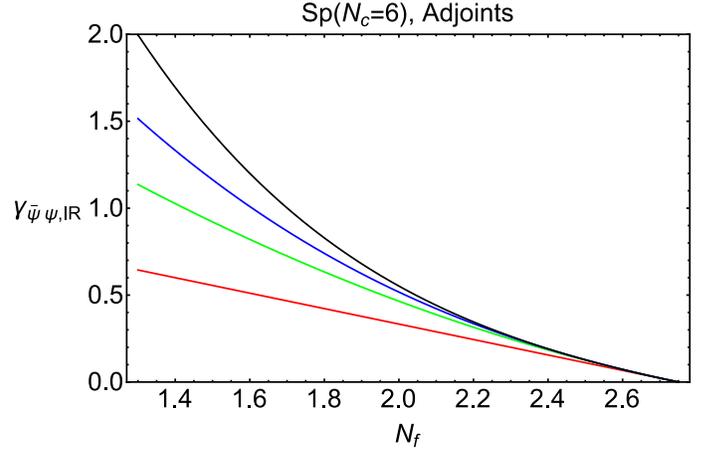}
  \end{center}
  \caption{Plot of $\gamma_{\bar\psi\psi,IR,A,\Delta_f^p}$ (labelled as
    $\gamma_{\bar\psi\psi,IR}$ on the vertical axis) for an Sp(6) gauge theory
    with fermions in the adjoint representation $R=A$, with $1 \le p \le 4$, as
    a function of $N_f \in I$. From bottom to top, the curves (with colors
    online) refer to $\gamma_{\bar\psi\psi,IR,A,\Delta_f}$ (red),
    $\gamma_{\bar\psi\psi,IR,A,\Delta_f^2}$ (green),
    $\gamma_{\bar\psi\psi,IR,A,\Delta_f^3}$ (blue), and
    $\gamma_{\bar\psi\psi,IR,A,\Delta_f^4}$ (black).}
\label{gamma_Sp6_adj_plot}
\end{figure}

For our two specific illustrative theories, SO(5) and Sp(6), the interval $I$
is the same and is given by Eq. (\ref{interval_adj}).  In
Figs. \ref{gamma_SO5_adj_plot} and \ref{gamma_Sp6_adj_plot} we show plots of
$\gamma_{\bar\psi\psi,IR,\Delta_f^p}$ with $1 \le p \le 4$ for this adjoint
case $R=A$, as a function of $N_f$ formally generalized to a real variable.
The curves are rather similar, as a consequence of the fact that $\kappa_{1,A}$
and $\kappa_{2,A}$ are the same and are independent of $N_c$, and, furthermore,
the differences between $\kappa_{j,A,{\rm SO}(5)}$ and 
$\kappa_{j,A,{\rm Sp}(6)}$ are small for $j=3,4$. 
As we found in our SU($N_c$) studies
\cite{gtr,dex,dexs,dexl}, the convergence of the $\Delta_f$ expansion is
slightly slower for $R=A$ than $R=F$, and this also tends to be true for the
other rank-2 tensor representations.  We find that, for both SO(5) and Sp(6),
as $N_f$, formally generalized to a real number, decreases in the interval $I$,
$\gamma_{\bar\psi\psi,IR}$ calculated to its highest order, $O(\Delta_f^4)$,
exceeds the conformality upper bound of 2 as $N_f$ reaches about $N_f \simeq
1.3$, before it decreases all the say to the lower end of this interval, at
$N_f=1.0625$.  This reduction in the non-Abelian Coulomb phase (conformal
window), relative to the full interval $I$ that we find here is similar to what
was observed for SU($N$) theories with higher representations in
\cite{rs_conformalwindow}.

In addition to the manifestly positive $\kappa_{1,A}$ and $\kappa_{2,A}$, we
find, by numerical evaluation, that $\kappa_{3,A}$ and $\kappa_{4,A}$ are
positive for all relevant $N_c$ for both types of gauge groups. 

Since the Lie algebras of SU(4) and SO(6) are isomorphic, it follows that
\beq
\kappa_{j,A,{\rm SO}(6)}=\kappa_{j,A,{\rm SU}(4)} \ .
\label{kappaj_su4adj_so6adj}
\eeq
This requirement serves as another check on our calculations. The check is
obviously satisfied for $\kappa_{1,A}$ and $\kappa_{2,A}$. 
Further, we obtain 
\beq
\kappa_{3,A,{\rm SO}(6)} = \kappa_{3,A,{\rm SU}(4)} = 
\frac{59209}{2^3 \cdot 3^{10}} = 0.125339
\label{kappa3_su4_adj}
\eeq
and
\beqs
\kappa_{4,A,{\rm SO}(6)} &=& \kappa_{4,A,{\rm SU}(4)} \cr\cr
&=& \frac{51983233}{2^7 \cdot 3^{14}} + \frac{3226}{3^{11}}\zeta_3 = 0.106800 
\cr\cr
&&
\label{kappa4_su4_adj}
\eeqs

In the LN limit, $\lim_{LN} \kappa_{j,A}$ is the same for SO($N_c$) and 
Sp($N_c$).  The coefficients $\kappa_{1,A}$ and $\kappa_{2,A}$ are evidently
independent of $N_c$. The values of $\kappa_{3,A}$ and 
$\kappa_{4,A}$ in the LN limit 
are (with numerical values given to the indicated precision) 
\beq
\lim_{N_c \to \infty} \kappa_{3,A} =\frac{61873}{2^3 \cdot 3^{10}} 
= 0.1309871
\label{kappa3_adj_Ncinf}
\eeq
and
\beq
\lim_{N_c \to \infty} \kappa_{4,A} = \frac{53389393}{2^7 \cdot 3^{14}} 
+ \frac{368}{3^{10}}\zeta_3 = 0.0946976 \ . 
\label{kappa4_adj_Ncinf}
\eeq
%


\subsection{ $R=S_2$ in SO($N_c$) and $R=A_2$ in Sp($N_c$) } 

It is convenient to give results for $R=S_2$ in SO($N_c$) and $R=A_2$ in
Sp($N_c$) together, since they are simply related by sign reversals in certain
terms. Recall that for SO($N_c$), $N_c$ must be $\ge 5$ if $R=S_2$ in order for
the theory to be asymptotically free.  In the following expressions, the upper
sign refers to $R=S_2$ in SO($N_c$) and the lower sign to $R=A_2$ in Sp($N_c$).
We will use a compact notation in which $T_2$ refers to these two respective
cases. From our general formulas we calculate
\beq
\kappa_{1,T_2} = \frac{4N_c(N_c\pm 2)}{(N_c \mp 2)(9N_c \mp 7)} \ , 
\label{kappa1_t2}
\eeq
\beq
\kappa_{2,T_2} = \frac{N_c(N_c \pm 2)^2(11N_c \mp 14)(93N_c \mp 10)}
{6(N_c \mp 2)^2(9N_c \mp 7)^3} \ , 
\label{kappa2_t2}
\eeq
\begin{widetext}
\beqs
\kappa_{3,T_2} &=& \frac{N_c(N_c \pm 2)^2}
{2^3 \cdot 3^3 (N_c \mp 2)^3(9N_c \mp 7)^5} \bigg [ 
\Big (1670571N_c^5 \mp 1075194N_c^4 -7188904N_c^3 \pm 14840368N_c^2 \cr\cr
&&+2671344N_c \mp 6795040 \Big ) 
\pm 2^{10} \cdot 33(9N_c \mp 7)(3N_c^3 \pm 23N_c^2 -38N_c \mp 56)\zeta_3 
\bigg ] \ , 
\label{kappa3_t2}
\eeqs
and 
\beqs
\kappa_{4,T_2} &=& \frac{N_c(N_c \pm 2)^3}
{2^7 \cdot 3^4 (N_c \mp 2)^4(9N_c \mp 7)^7} \, 
\bigg [ \Big ( 4324540833N_c^7 \mp 6239517858N_c^6 -9953927772N_c^5 
\pm 61550306040N_c^4 \cr\cr
&&-90479597392N_c^3 \mp 24158962016N_c^2 + 61198146240N_c 
\mp 11095638400 \Big ) \cr\cr
&&+2^{10} (9N_c \mp 7)\Big (33534N_c^6 \pm 743769N_c^5+4721805N_c^4
\mp 16060070N_c^3-5795540N_c^2 \pm 16964328N_c+3786048 \Big )\zeta_3 \cr\cr
&& \mp 2^{14} \cdot 275(N_c \mp 2)(9N_c \mp 7)^2
(15N_c^3 \pm 139N_c^2 +234N_c \pm 120)\zeta_5 \ \bigg ] \ . 
\label{kappa4_t2}
\eeqs
\end{widetext} 

\begin{figure}
  \begin{center}
    \includegraphics[height=6cm]{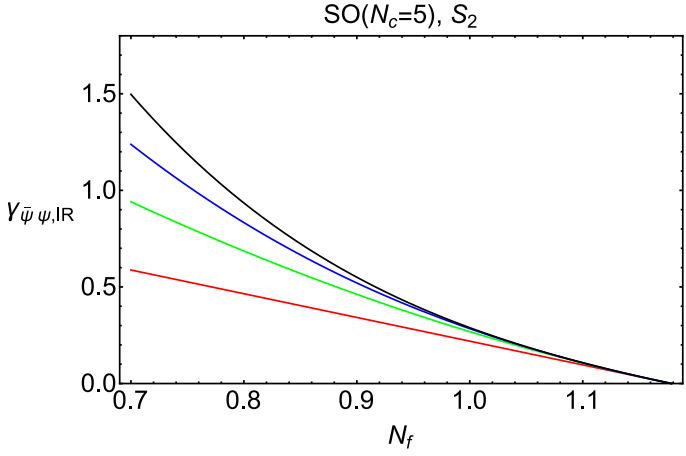}
  \end{center}
\caption{Plot of $\gamma_{\bar\psi\psi,IR,S_2,\Delta_f^p}$ (labelled as
$\gamma_{\bar\psi\psi,IR}$ on the vertical axis)
for an SO(5) gauge theory with fermions in the $S_2$ representation, with
$1 \le p \le 4$, as a function of $N_f \in I$. From bottom to top,
the curves (with colors online) refer to
$\gamma_{\bar\psi\psi,IR,S_2,\Delta_f}$ (red),
$\gamma_{\bar\psi\psi,IR,S_2,\Delta_f^2}$ (green),
$\gamma_{\bar\psi\psi,IR,S_2,\Delta_f^3}$ (blue), and
$\gamma_{\bar\psi\psi,IR,S_2,\Delta_f^4}$ (black).}
\label{gamma_SO5_sym_plot}
\end{figure}

\begin{figure}
  \begin{center}
    \includegraphics[height=6cm]{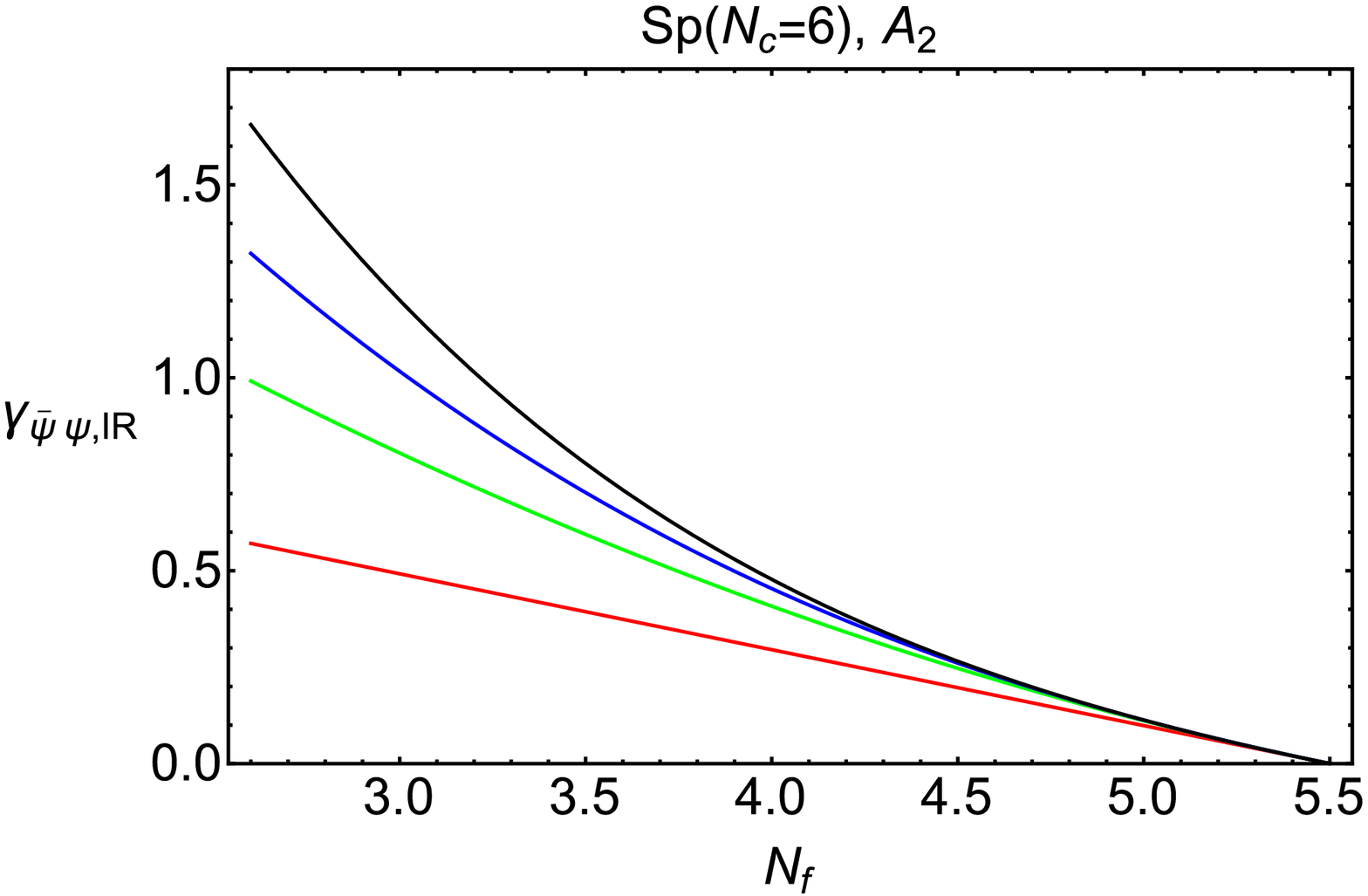}
  \end{center}
\caption{Plot of $\gamma_{\bar\psi\psi,IR,A_2,\Delta_f^p}$ (labelled as
$\gamma_{\bar\psi\psi,IR}$ on the vertical axis)
for an Sp(6) gauge theory with fermions in the $A_2$ representation, with
$1 \le p \le 4$, as a function of $N_f \in I$. From bottom to top,
the curves (with colors online) refer to
$\gamma_{\bar\psi\psi,IR,A_2,\Delta_f}$ (red),
$\gamma_{\bar\psi\psi,IR,A_2,\Delta_f^2}$ (green),
$\gamma_{\bar\psi\psi,IR,A_2,\Delta_f^3}$ (blue), and
$\gamma_{\bar\psi\psi,IR,A_2,\Delta_f^4}$ (black).}
\label{gamma_Sp6_asym_plot}
\end{figure}

We next apply these results for our two specific illustrative theories, SO(5)
and Sp(6).  In the SO(5) theory with $R=S_2$, $N_{u,{\rm SO}(5),S_2}=33/28
=1.1786$ and $N_{\ell,{\rm SO}(5),S_2}=51/140 = 0.3643$, while in the Sp(6)
theory with $R=A_2$, $N_{u,{\rm Sp}(6),S_2}=5.5$ and $N_{\ell,{\rm
    Sp}(6),S_2}=68/29=2.345$.  In Figs.  \ref{gamma_SO5_sym_plot} and
\ref{gamma_Sp6_asym_plot} we show plots of
$\gamma_{\bar\psi\psi,IR,R,\Delta_f^p}$ with $1 \le p \le 4$ for SO(5) with
$R=S_2$ and for Sp(6) with $R=A_2$, respectively, with $N_f$ formally
generalized to a real number. We see that in the SO(5) theory, as $N_f$
decreases in the interval $I$, $\gamma_{\bar\psi\psi,S_2,IR}$ calculated to its
highest order, $O(\Delta_f^4)$, exceeds the conformality upper bound $N_f \le
2$ reaches about $N_f \simeq 0.7$, well above the lower end of $I$ at 0.3643.
In the Sp(6) theory, as $N_f$ decreases in the interval $I$,
$\gamma_{\bar\psi\psi,A_2,IR}$ calculated to its highest order,
$O(\Delta_f^4)$, exceeds the conformality upper bound $N_f \le 2$ reaches about
$N_f \simeq 2.4$, close to the lower end of $I$ at 2.345.

In addition to the manifestly positive $\kappa_{1,T_2}$ and $\kappa_{2,T_2}$, 
we find, by numerical evaluation, that $\kappa_{3,T_2}$ and $\kappa_{4,T_2}$ 
are positive for all relevant $N_c$ in these SO($N_c$) and Sp($N_c$) theories. 

These coefficients have the same LN limits as the $\kappa_{j,A}$
\beq
\lim_{N_c \to \infty} \kappa_{j,T_2} = \lim_{N_c \to \infty} \kappa_{j,A} \ . 
\label{kappa_sym_kappa_adj}
\eeq
%


\section{Calculation of $\beta'_{IR}$ to $O(\Delta_f^5)$ Order} 
\label{betaprime_section}

\subsection{$R=F$} 

For the coefficients $d_j$, we recall first that $d_1=0$ for all $G$ and $R$. 
As was true of the $\kappa_{j,R}$ coefficients, the $d_{j,R}$ coefficients for 
SO($N_c$) and Sp($N_c$) are simply related to each other with sign reversals in
various terms, and hence it is natural to present them together.  
Concerning the signs of these coefficients, our general expressions in
\cite{dex} for $d_2$ and $d_3$ show that they are positive for arbitrary $G$
and $R$:
\beq
d_2 = \frac{2^5 T_f^2}{3^2 C_A (7C_A+11C_f)} \ ,
\label{d2}
\eeq
and
\beq
d_3 = \frac{2^7 T_f^3(5C_A+3C_f)}{3^3 C_A^2 (7C_A+11C_f)^2} \ . 
\label{d3}
\eeq
Since our general expressions for $d_4$ in \cite{dex} and for $d_5$ in 
\cite{dexs,dexl} contain negative terms, it is necessary to investigate the
signs of these terms as a function of $G$, $R$, and $N_c$. 

For the fundamental representation, we obtain the following results, where, as
before, the upper and lower signs refer to SO($N_c$) and Sp($N_c$),
respectively:
\beq
d_{2,F} = \frac{2^6}{3^2(N_c \mp 2)(25N_c \mp 39)} \ , 
\label{d2_fund}
\eeq
\beq
d_{3,F} = \frac{2^8(13N_c \mp 23)}{3^3(N_c \mp 2)^2(25N_c \mp 39)^2} \ , 
\label{d3_fund}
\eeq
\begin{widetext}
\beqs
d_{4,F} &=& \frac{2^8}{3^5(N_c \mp 2)^3(25N_c \mp 39)^5} \, \bigg [
\Big ( 366782N_c^4 \mp 2269256N_c^3+5506308N_c^2 \mp 6383412N_c +2994975 
\Big ) \cr\cr
&&-2^5 \cdot 33(N_c \mp 3)(25N_c\mp 39)(25N_c^2 \mp 65N_c+94 \Big )\zeta_3 
\ \bigg ] \ , 
\label{d4_fund}
\eeqs
and
\beqs
d_{5,F} &=& \frac{2^{10}}{3^6(N_c \mp 2)^4(25N_c \mp 39)^7} \, \bigg [
\Big ( -298194551N_c^6 \pm 3084573642N_c^5 -13173836397N_c^4 \cr\cr
&&\pm 29649471936N_c^3 -37042033788N_c^2 \pm 24377774904N_c -6624643320 \Big ) 
\cr\cr
&&-2^5 (25N_c\mp 39)(529125N_c^5 \mp 4349794N_c^4 + 14556219N_c^3
\mp 23420126N_c^2 + 15005784N_c \mp 467496 \Big )\zeta_3 \cr\cr
&& +2^7 \cdot 55 (N_c \mp 2)(N_c \mp 3)(25N_c \mp 39)^2(120N_c^2 \mp 229N_c
+ 511)\zeta_5 \ \bigg ] \ . 
\eeqs
\end{widetext}
In addition to the manifestly positive $d_2$ and $d_3$, for SO($N_c$), we 
find that $d_{4,F}$ is positive if $N_c=3$, but decreases through zero and is
negative for large $N_c$, while $d_{5,F}$ is
negative for the relevant range $N_c$.  For Sp($N_c$), we find that 
both $d_{4,F}$ and $d_{5,F}$ are negative in the relevant range of (even)
$N_c$. 

As $N_c \to \infty$, the $d_{j,F} \propto 1/N_c^j + O(1/N_c^{j+1})$, and hence
the finite coefficients for the scheme-independent
expansion of $\beta'_{IR}$ in this limit are 
\beq
\hat d_{j,F} = \lim_{N_c \to \infty} N_c^j d_{j,F} \ . 
\label{dhatj}
\eeq
These limiting values are the same for SO($N_c$) and Sp($N_c$). 
From our results above, we calculate
\beq
\hat d_{2,F} = \frac{2^6}{3^2 \cdot 5^2} = 0.284444 \ , 
\label{dhat2}
\eeq
\beq
\hat d_{3,F} = \frac{2^8 \cdot 13}{3^3 \cdot 5^4} = 0.197215 \ , 
\label{dhat3}
\eeq
\beqs
\hat d_{4,F} &=& \frac{2^9 \cdot 183391}{3^5 \cdot 5^{10}} 
-\frac{2^{13} \cdot 11}{3^4 \cdot 5^6}\zeta_3 \cr\cr
&=& -0.0460182 \ , 
\label{dhat4}
\eeqs
and
\beqs
\hat d_{5,F} &=& -\frac{2^{10} \cdot 298194551}{3^6 \cdot 5^{14}} 
-\frac{2^{15} \cdot 1411}{3^5 \cdot 5^9}\zeta_3 
+ \frac{2^{20} \cdot 11}{3^5 \cdot 5^8}\zeta_5 \cr\cr
&=& -0.0597277 \ . 
\label{dhat5}
\eeqs
%


\subsection{$R=A$} 

As discussed above, for the SO($N_c$) and Sp($N_c$) theories with $R=A$, the
adjoint representation, only one value of $N_f$ is allowed by asymptotic 
freedom and lies in the interval $I$, namely $N_f=2$. We calculate the 
following results for the $d_{j,A}$, with $N_c$ kept in as a formal 
variable (and with numerical values given to the indicated precision) 
\beq
d_{2,A}=\Big ( \frac{2}{3} \Big )^4 = 0.197531 \ , 
\label{d2_adj}
\eeq
\beq
d_{3,A}=\frac{2^8}{3^7} = 0.117055 \ , 
\label{d3_adj}
\eeq
\begin{widetext}
\beqs
d_{4,A} &=&\frac{1}{2^2 \cdot 3^{12} (N_c \mp 2)^3} \, \Big (
46871N_c^3 \mp 302538N_c^2+860820N_c \mp 1056952 \Big ) \ , 
\label{d4_adj}
\eeqs
and
\beqs
d_{5,A} &=& \frac{1}{2^3 \cdot 3^{16}(N_c \mp 2)^3} \, \bigg [
\Big ( -7141205N_c^3 \pm 43403934N_c^2 -93488316N_c \pm 74944168 \Big ) \cr\cr
&&+ \Big (3566592N_c^3 \pm 3718656N_c^2 - 308855808N_c \pm 775249920 
\Big )\zeta_3
\bigg ] \ . 
\label{d5_adj}
\eeqs
\end{widetext}

The $N_c \to \infty$ limits of $d_{j,A}$ are the same for SO($N_c$) and 
Sp($N_c$).  We have 
\beq
\lim_{N_c \to \infty} d_{4,A} = \frac{46871}{2^2 \cdot 3^{12}} = 
2.204901 \times 10^{-2}  
\label{d4_adj_Ncinf}
\eeq
and
\beqs
\lim_{N_c \to \infty} d_{5,A} &=& -\frac{7141205}{2^3 \cdot 3^{16}} 
+ \frac{2^7 \cdot 43}{3^{12}} \zeta_3 \cr\cr
&=& -(0.8287386 \times 10^{-2}) \ . 
\label{d5_adj_Ncinf}
\eeqs
In addition to the manifestly positive $d_{2,A}$ and $d_{3,A}$, we find that
for SO($N_c$), in the relevant range of $N_c$, $d_{4,A}$ is positive, while 
$d_{5,A}$ is negative. For Sp($N_c$), $d_{4,A}$ is manifestly positive, 
and we find that $d_{5,A}$ is negative. 


\subsection{ $R=S_2$ in SO($N_c$) and $R=A_2$ in Sp($N_c$) } 

As before, we present our results for $R=S_2$ in SO($N_c$) and $R=A_2$ in 
Sp($N_c$) together, since they are simply related by sign reversals in certain
terms. Recall that for SO($N_c$), $N_c$ must be $\ge 5$ if $R=S_2$ in order for
the theory to be asymptotically free. 
In the following expressions, the upper sign refers to 
$R=S_2$ in SO($N_c$) and the lower sign to $R=A_2$ in Sp($N_c$). 
We again use the compact notation in which $T_2$ refers to these two respective
cases. From our general formulas we calculate 
\beq
d_{2,T_2} = \frac{2^4(N_c \pm 2)^2}{3^2(N_c \mp 2)(9N_c \mp 7)} \ , 
\label{d2_t2}
\eeq
\beq
d_{3,T_2} = \frac{2^6(N_c \pm 2)^3(4N_c \mp 5)}
{3^3 (N_c \mp 2)^2(9N_c \mp 7)^2} \ , 
\label{d3_t2}
\eeq
\begin{widetext}
\beqs
d_{4,T_2} &=& \frac{(N_c \pm 2)^3}{2^2 \cdot 3^5(N_c \mp 2)^3(9N_c \mp 7)^5} \,
\bigg [ \Big (1265517N_c^5 \mp 618894N_c^4 +3021512N_c^3 \mp 10811760N_c^2 
\cr\cr
&& -16432368N_c \pm 16806048 \Big ) 
\pm 2^{12} \cdot 33(9N_c \mp 7)(3N_c^3 \mp 15N_c^2 + N_c \pm 42)\zeta_3 \ 
\bigg ] \ , 
\label{d4_t2}
\eeqs
and 
\beqs
d_{5,T_2} &=& \frac{(N_c \pm 2)^4}{2^3 \cdot 3^6(N_c \mp 2)^4(9N_c \mp 7)^7} \,
\bigg [ \Big ( -578437605N_c^7 \pm 3437217450N_c^6 -6404128380N_c^5 
\mp 13828926056N_c^4 \cr\cr
&&+52499838288N_c^3 \mp 21845334432N_c^2 -14381806656N_c \pm 6247244416 
\Big ) \cr\cr
&&+2^9 (9N_c \mp 7)\Big (62694N_c^6 \pm 61965N_c^5 -6430023N_c^4 
\pm 11443586N_c^3 + 10920884N_c^2 \mp 16105176N_c -1862112 \Big )\zeta_3 \cr\cr
&&\mp 2^{13} \cdot 55 (N_c \mp 2)(N_c \mp 9)(9N_c \mp 7)^2
(87N_c^2 \pm 178N_c + 48)\zeta_5 \ \bigg ] \ . 
\label{d5_t2}
\eeqs
\end{widetext} 
Concerning signs, in addition to the manifestly positive $d_{2,T_2}$ and
$d_{3,T_2}$, we find that for SO($N_c$) with $N_c \ge 5$, $d_{4,S_2} > 0$ and
$d_{5,S_2} < 0$, while for Sp($N_c$), $d_{4,A_2} < 0$ if $N_c=4$, $d_{4,A_2} >
0$ if $N_c \ge 6$, and $d_{5,A_2} < 0$ for all $N_c \ge 4$. We further note
that
\beq
\lim_{N_c \to \infty} d_{j,T_2} = \lim_{N_c \to \infty} d_{j,A} \ . 
\label{dj_adj_symasym_Ncinf}
\eeq
%


\section{Conclusions}
\label{conclusions_section}

In this paper we have used our general calculations in \cite{gtr,gsi,dex,dexl}
to obtain scheme-independent results for the anomalous dimension,
$\gamma_{\bar\psi\psi,IR}$, and the derivative of the beta function,
$\beta'_{IR}$, at an infrared fixed point of the renormalization group in the
non-Abelian Coulomb phase of vectorial, asymptotically free SO($N_c$) and (with
even $N_c$) Sp($N_c$) gauge theories with fermions in several different
irreducible representations, namely fundamental, adjoint, and rank-2 symmetric
and antisymmetric tensor. We calculate $\gamma_{\bar\psi\psi,IR}$ to
$O(\Delta_f^4)$ and $\beta'_{IR}$ to $O(\Delta_f^5)$, where $\Delta_f$ is the
expansion parameter defined in Eq. (\ref{deltaf}). These are the highest orders
to which these quantities have been calculated for these theories. Our present
results extend our earlier ones for the case of SU($N_c$) gauge theories in
\cite{gtr,gsi,dex,dexs,dexl} to these other two types of gauge groups. 

An important question that we address and answer is whether the coefficients
$\kappa_j$ in the expansion (\ref{gamma_delta_series}) are positive for
SO($N_c$) and Sp($N_c$) with all of the representations that we consider, just
as we found earlier for SU($N_c$).  We find that the answer is affirmative.
Our finding yields two monotonicity results for these SO($N_c$) and Sp($N_c$)
groups and representations, namely that (i)
$\gamma_{\bar\psi\psi,IR,\Delta_f^p}$ increases monotonically as $N_f$
decreases from $N_u$ in the non-Abelian Coulomb phase; (ii) for a fixed $N_f$
in the NACP, $\gamma_{\bar\psi\psi,IR,\Delta_f^p}$ increases monotonically with
$p$. Our results in this paper provide further support for our conjecture that,
in addition to the manifestly positive $\kappa_1$ and $\kappa_2$, the
$\kappa_j$ for $j \ge 3$ are positive for a vectorial asymptotically free gauge
theory with a general (simple) gauge group $G$ and fermion representations $R$
that we have considered.  In turn, this conjecture implies several monotonicity
properties, namely the generalizations of (i) and (ii) to arbitrary $p$ and
thus the property that the quantity $\gamma_{\bar\psi\psi,IR}$ defined by the
infinite series (\ref{gamma_delta_series}), increases
monotonically with decreasing $N_f$ in the non-Abelian Coulomb phase.  Using
this property in conjunction with the upper bound on $\gamma_{\bar\psi\psi,IR}$
in a conformally invariant theory, and the assumption that this bound is
saturated at the lower end of the NACP (as it is in the exact results for an
${\cal N}=1$ supersymmetric gauge theory), we have given estimates of the lower
end of this non-Abelian Coulomb phase for illustrative theories of these
types. 


\begin{acknowledgments}

The research of T.A.R. and R.S. was supported in part by the Danish National
Research Foundation grant DNRF90 to CP$^3$-Origins at SDU and by the
U.S. National Science Foundation Grant NSF-PHY-16-1620628, respectively.  

\end{acknowledgments}


\begin{appendix}

\section{Some Group-Theoretic Quantities}
\label{group_invariants}

In this appendix we discuss some group-theoretic quantities that enter in our
calculations. As in the text, we denote the gauge group as $G$.  The generators
of the Lie algebra of this group, in the representation $R$, are 
denoted $T^a_R$, with $1 \le a \le d_A$. The generators satisfy the Lie algebra
\beq
[T^a_R,T^b_R]=if^{abc} T^c_R \ ,
\label{algebra}
\eeq
where the $f^{abc}$ are the associated structure constants of this Lie algebra.
Here and elsewhere a sum over repeated indices is understood. We denote the
dimension of a given representation $R$ as $d_R = {\rm dim}(R)$. In particular,
as in the text, we denote the adjoint representation by $A$, with the dimension
$d_A$ equal to the number of generators of the group, i.e., the order of the
group. (The dimension $d_A$ should not be confused with the tensors
$d_A^{abcd}$.)  The normalization of the generators is given by the trace in
the representation $R$,
\beq
{\rm Tr}_R(T^a_R T^b_R) = T(R)\delta_{ab} \ .
\label{trace}
\eeq
The quadratic Casimir invariant $C_2(R)$ is given by
\beq
T^a_RT^a_R = C_2(R) I \ , 
\label{c2}
\eeq
where $I$ is the $d_R \times d_R$ identity matrix. For a fermion $f$
transforming according to a representation $R$, we often use the equivalent
compact notation $T_f \equiv T(R)$ and $C_f \equiv C_2(R)$. We also use the
notation $C_A \equiv C_2(A) \equiv C_2(G)$.  The invariants $T(R)$ and $C_2(R)$
are related according to
\beq
C_2(R)d_R = T(R)d_A \ . 
\label{c2trel}
\eeq

A remark on the normalization of the generators is in order.  As was noted in
\cite{b4,invariants}, although the normalization $T(F)=1/2$, where $F$ is the
fundamental representation, is standard for the trace in Eq. (\ref{trace}) for
SU($N$), two normalizations are widely used for this normalization for SO($N$)
and Sp($N$) groups. As indicated, our normalization is $T(F)=1$ for SO($N$) and
$T(F)=1/2$ for Sp($N$).  If one multiplies $T(R)$ by a factor $\rho$, this is
equivalent to multiplying the generators and structure constants by
$\sqrt{\rho}$ and the quadratic Casimir invariant $C_2(R)$ by $\rho$.  In the
covariant derivative $D_\mu = \partial_\mu \cdot 1 - g{\vec T} \cdot {\vec
  A}_\mu$, where $A^a_\mu$ is the gauge field, a rescaling of the generators by
$\sqrt{\rho}$ means that $g$ is rescaling by $1/\sqrt{\rho}$, with the gauge
field continuing to have canonical normalization.  Physical quantities such as
$N_u$, $N_\ell$, $\gamma_{\bar\psi\psi,IR}$, and $\beta'_{IR}$ are independent
of this normalization convention with $\rho$.  This can be seen from
Eqs. (\ref{nfb1z}), (\ref{nfb2z}), and the explicit expressions that we have
given in our earlier works \cite{gtr,gsi,dex,dexs,dexl} for the coefficients
$\kappa_j$ and $d_j$.  For example, in the expressions $\kappa_2 =
8C_fT_f/[C_A(7C_A+11C_f)]$ and $d_2 = 32T_f^2/[9C_A(7C_A+11C_f)]$, both the
numerator and denominator scale like $\rho^2$, so this normalization factor
cancels, and similarly for other $\kappa_j$ and $d_j$.

In this appendix we will, for generality, consider the three types of gauge
groups SU($N$), SO($N$), and Sp($N$). As noted before, the correspondence
between the mathematical notation for the Cartan series of Lie algebras and our
notation used here is $A_n = {\rm SU}(n+1)$, $B_n = {\rm SO}(2n+1)$, $C_n =
{\rm Sp}(2n)$, and $D_n = SO(2n)$.  One may recall some basic properties of
these Lie groups and their associated Lie algebras (see, e.g.,
\cite{grouptheoryrefs}-\cite{patera}).  Concerning representations, SU(2) has
only real representations, while SU($N$) with $N \ge 3$ has complex
representations. Sp($N$) ($N$ even) and SO($N$) with odd $N$ have only real
representations, while SO($N$) with even $N$ also have both real and complex
representations. Concerning the values of $N_f$, we note that for a real
representation, one could consider half-integral $N_f$, corresponding to a
Majorana fermion.  However, this would entail a global Witten anomaly
associated with the homotopy group $\pi_4(G)$ in the case $G={\rm SO}(N)$ with
$N=3, \ 4, \ 5$, and for all Sp($N$) (while $\pi_4({\rm SO}(N))=\emptyset$ for
$N \ge 6$ \cite{finkelstein}.) Hence, we restrict to integer $N_f$, i.e., Dirac
fermions.

In Tables \ref{invariants_table} we list the dimensions and
quadratic group invariants for SU($N$), SO($N$), and Sp($N$) groups with the
various representations considered in the text \cite{patera}. 
The results for SU($N$) are
well-known, but some remarks are in order for SO($N$) and Sp($N$).
An element $O$ of SO($N$) satisfies $OO^T=1$.  Starting with a 2-index tensor
$\psi = \psi^{ij}$ of SO($N$), we can form symmetric and antisymmetric
quantities in the obvious way by taking sums and differences of $\psi$ and
$\psi^T$. However, to form the irreducible symmetric representation of SO($N$),
$S_2$, it is necessary to remove the trace, so we write
\beqs
\psi &=& \frac{1}{2}(\psi+\psi^T) - {\rm Tr}(\psi) \cdot 1 \cr\cr
     &+& \frac{1}{2}(\psi-\psi^T) \cr\cr
     &+& {\rm Tr}(\psi) \cdot 1 \ , 
\label{psi_so}
\eeqs
where here $1$ is the $N \times N$ identity matrix. 
The quantities in the first and second lines of Eq. (\ref{psi_so}) form the 
(traceless) $S_2$ and $A_2$ representations of 
SO($N$) (the latter being automatically traceless), while the quantity in the
third line is a singlet.  The dimensions of the $S_2$ and $A_2$ 
representations are therefore 
\beq
d_{S_2,{\rm SO}(N)} = \frac{N(N+1)}{2}-1 = \frac{(N-1)(N+2)}{2}
\label{dim_sym_so}
\eeq
and $d_{A_2,{\rm SO}(N)} = N(N-1)/2 = d_{A,{\rm SO}(N)}$, as listed in
the table. 

An element $S$ of Sp($N$) satisfies $SES^T=E$, with $E$ the antisymmetric
$N \times N$ matrix
\beq
E = \left( \begin{array}{cc}
    0  & 1 \\
    -1 & 0 \end{array} \right ) \ , 
\label{ematrix}
\eeq
where here the symbols $0$ and $1$ denote $N/2 \times N/2$ submatrices. 
We can thus write 
\beqs
\psi &=& \frac{1}{2}(\psi + \psi^T) \cr\cr
     &+& \frac{1}{2}(\psi - \psi^T) - {\rm Tr}(\psi)E \cr\cr
     &+& {\rm Tr}(\psi)E \ , 
\label{psi_sp}
\eeqs
The quantities in the first and second lines of Eq. (\ref{psi_sp}) form the 
$S_2$ and $A_2$ representations of Sp($N$), while the 
third line is a singlet.  The dimensions of the $S_2$ and $A_2$ 
representations are therefore 
$d_{S_2,{\rm Sp}(N)} = N(N+1)/2 = d_{A,{\rm Sp}(N)}$, and 
\beq
d_{A_2,{\rm Sp}(N)} = \frac{N(N-1)}{2}-1=\frac{(N+1)(N-2)}{2} \ , 
\label{dim_asym_sp}
\eeq
as listed in the table. We remark that the 
expressions for $T(R)$ and $C_2(R)$ for Sp($N$) are simply 
related to those for SO($N$) by a factor of 1/2 and sign reversals of certain
terms.

At the four-loop and five-loop level, new types of group-theoretic invariants
appear in the coefficients for the beta function and anomalous dimension
$\gamma_{\bar\psi\psi,IR}$, namely the four-index quantities
$d^{abcd}_R$.  For a given representation $R$ of $G$, 
\beqs
d^{abcd}_R &=& 
\frac{1}{3!} {\rm Tr}_R\Big [ T_a(T_bT_cT_d + T_bT_dT_c + T_cT_bT_d \cr\cr
&+& T_cT_dT_b + T_dT_bT_c + T_dT_cT_b) \Big ] \ . 
\label{d_abcd}
\eeqs
From Eq. (\ref{d_abcd}), it is evident that $d^{abcd}_R$ is a totally symmetric
function of the group indices $a,b,c,d$. One can express this as
\beqs
d_R^{abcd} &=& I_{4,R}d^{abcd} + \Big ( \frac{T(R)}{d_A+2}\Big )
\Big ( C_2(R) - \frac{1}{6}C_A\Big ) \times \cr\cr
&\times& 
(\delta^{ab}\delta^{cd} + \delta^{ac}\delta^{bd} + \delta^{ad}\delta^{bc}) \ , 
\label{d_abcd_form}
\eeqs
where $d^{abcd}$ is traceless (i.e., $\delta_{ab}d^{abcd}=0$, etc.), $I_{4,R}$
is a quartic group invariant (index) \cite{okubo,patera}, and $d_A$ is the
dimension of the adjoint representation, i.e., the number of generators of the
Lie algebra of $G$. The traceless tensor $d^{abcd}$ depends only on the group
$G$, but not on the representation $R$.  The quartic indices $I_{4,R}$ are
listed for the relevant representations in Table \ref{I4}. The quantities that
appear in the coefficients that we calculate involve products of these
$d_R^{abcd}$ of the form $d_R^{abcd}d_{R'}^{abcd}$, summed over the group
indices $a, \ b, \ c, \ d$.  These can be written as
\beqs
&& d^{abcd}_R d^{abcd}_{R'} = I_{4,R}I_{4,R'}d^{abcd}d^{abcd} + \cr\cr
&&+ \Big (\frac{3d_A}{d_A+2} \Big )T(R)T(R')\Big ( C_R - \frac{1}{6}C_A \Big )
\Big ( C_{R'} - \frac{1}{6}C_A \Big ) \ . \cr\cr
&& 
\label{ddform}
\eeqs

One has, for the quartic Casimir invariants that depend only on $G$, the
results \cite{b4,invariants}
\beq
{\rm SU}(N): \quad d^{abcd}d^{abcd} = \frac{d_A(d_A-3)(d_A-8)}{96(d_A+2)} \ , 
\label{dd_sun}
\eeq
\beq
{\rm SO}(N): \quad d^{abcd}d^{abcd} = \frac{d_A(d_A-1)(d_A-3)}{12(d_A+2)} \ , 
\label{dd_son}
\eeq
and
\beq
{\rm Sp}(N): \quad d^{abcd}d^{abcd} = \frac{d_A(d_A-1)(d_A-3)}{192(d_A+2)} \ , 
\label{dd_spn}
\eeq
so that $d^{abcd}d^{abcd}$ for Sp($N$) is formally 1/16 times the corresponding
quantity for SO($N$) (with different $d_A$ understood).  Note that
$d^{abcd}d^{abcd}=0$ for SU(2), SO(3), and Sp(2), since the dimension of the
adjoint representation in all three cases is $d_A=3$.  This is in agreement
with the isomorphisms ${\rm SU}(2) \cong {\rm Sp}(2)$ and ${\rm SU}(2)
\cong {\rm SO}(3)$.  (These may be considered to refer to the Lie algebras;
for our purposes, we do not have to distinguish between local and global
isomorphisms.) Note also that $d^{abcd}d^{abcd}=0$ for SU(3), since $d_A=8$ for
SU(3). 

We list the resultant values of $d_R^{abcd}d_{R'}^{abcd}$ in Tables
\ref{dd}. As is evident from these tables, the expressions for the
$d_R^{abcd}d_{R'}^{abcd}$ for Sp($N$) are simply related to those for SO($N$)
by an overall factor of $1/16$ and sign reversals of certain
coefficients.  Our results for SU($N$) agree with the corresponding entries in
Table II in \cite{ps}; however, our results for $d^{abcd}_Rd^{abcd}_R$ and
$d^{abcd}_Rd^{abcd}_{A}$ differ from those given in Table II of \cite{ps} for
SO($N$) and Sp($N$) \cite{comm}.  We have performed several checks on the
correctness of our results:

\begin{enumerate} 

\item Since ${\rm SU}(4) \cong {\rm SO}(6)$, the coefficients $\kappa_j,\
  j=1,\dots,4$ and $d_j,\ j=1\dots,5$ calculated for ${\rm SU} (4)$ must agree
  with their counterparts for SO(6) when the matter representations are
  equivalent. We have checked that this is satisfied in a number of
  cases. Specifically, this must hold for (i) the 20-dimensional $S_2$
  representation of SO(6) and the real 20-dimensional representation of $SU(6)$
  with Dynkin label (0,2,0); (ii) the fundamental 6-dimensional representation
  of SO(6) and the 6-dimensional $A_2$ representation of SU(4); and (iii) the
  adjoint representation of both SU(4) and SO(6). The group invariants for the
  real 20-dimensional representation of SU(4) with Dynkin label (0,2,0) we 
  have used are $T_f = 8$ and $C_f = 6$ \cite{perelman_popov}.
  
\item Since the adjoint representation of SU(2) is equivalent to the adjoint as
  well as the fundamental representation of SO(3), it follows that the
  corresponding coefficients $\kappa_j,\ j=1,\dots,4$ and $d_{j},\ j=1\dots,5$
  should be equal, and we have verified that this is the case.

\item Since ${\rm SU}(2) \cong {\rm Sp}(2)$, it follows that the expressions
  for $\kappa_j$ and $d_j$ should be the same for our representations $R$ 
  for these two groups, and they are.

\item The isomorphism SO(5) $\cong$ Sp(4) \cite{pss} yields a further check on
  our results.  The invariants for the adjoint representations of these groups
  must be equal and they are. Further, the fundamental representation of SO(5)
  has the same dimension as the $A_2$ representation of Sp(4), and these yield
  the same $\kappa_j$ and $d_j$ values, which provides a check on our
  expressions for the $A_2$ representation of Sp($N$).

\end{enumerate} 

\end{appendix}



\newpage

\begin{widetext} 
\begin{table}
  \caption{\footnotesize{Values of various group invariants
      for the groups SU($N$), SO($N$), and (with even $N$) Sp($N$)  
      and (irreducible) fermion representations $R$ equal to fundamental 
      ($F$), adjoint ($A$),
      and rank-2 symmetric ($S_2$) and antisymmetric ($A_2$) tensor.
      We take $N \ge 2$ for SU($N$), $N \ge 3$ for SO($N$), and even $N \ge 2$
      for Sp($N$). Here, $d_R$ denotes the dimension of the representation $R$.
      For a fermion $f$ in the representation $R$, the equivalent
      compact notation $T_f \equiv T(R)$ and $C_f \equiv C_2(R)$ is used in the text.}}
\begin{center}
\begin{tabular}{|c||c|c|c|} \hline\hline
 \multicolumn{4}{|c|}{${\rm SU}(N), \ N \geq 2$} \\
\hline
 $R$ & $d_R$ & $T(R)$ & $C_2(R)$ \\
 \hline\hline
 $F$  & $N$  &  $\frac{1}{2}$  & $\frac{N^2-1}{2N}$  \\
\hline
 $A$ & $N^2-1$  &  $N$  & $N$   \\
\hline
 $S_2$ & $\frac{N(N+1)}{2}$ & $\frac{N+2}{2}$ &  $\frac{(N-1)(N+2)}{N}$  \\
 \hline
 $A_2$ & $\frac{N(N-1)}{2}$ & $\frac{N-2}{2}$ &  $\frac{(N+1)(N-2)}{N}$  \\
\hline\hline
\end{tabular}

\bigskip

\qquad 
\begin{tabular}{|c||c|c|c|} \hline\hline
 \multicolumn{4}{|c|}{${\rm SO}(N), \ N \geq 3$} \\
\hline
 $R$ & $d_R$ & $T(R)$ & $C_2(R)$  \\
\hline\hline
 $F$  & $N$  &  $1$  & $\frac{N-1}{2}$     \\
\hline
 $A$ & $\frac{N(N-1)}{2}$ & $N-2$ & $N-2$    \\
\hline
 $S_2$ & $\frac{(N-1)(N+2)}{2}$ & $N+2$ & $N$   \\
\hline\hline
\end{tabular}

\bigskip

\begin{tabular}{|c||c|c|c|} \hline\hline
 \multicolumn{4}{|c|}{${\rm Sp}(N), \ N \geq 2$} \\
 \hline
 $R$ & $d_R$ & $T(R)$ & $C_2(R)$ \\
 \hline \hline
 $F$  & $N$  &  $\frac{1}{2}$  & $\frac{N+1}{4}$    \\
\hline
 $A$ & $\frac{N(N+1)}{2}$  &  $\frac{N+2}{2}$  & $\frac{N+2}{2}$
    \\
\hline
 $A_2$ & $\frac{(N+1)(N-2)}{2}$  &  $\frac{N-2}{2}$  & $\frac{N}{2}$
  \\
\hline\hline
\end{tabular}
\end{center}
\label{invariants_table}
\end{table}


\begin{table}
  \caption{\footnotesize{Values of $I_{4,f}$ indices 
   for the groups SU($N$), SO($N$), and (with even $N$), Sp($N$) and
   and fermion representations $R$ equal to fundamental ($F$), adjoint ($A$),
   and rank-2 symmetric ($S_2$) and antisymmetric ($A_2$) tensor.
   $S_2$ (symmetric rank-2 tensor).}}
\begin{center}
\begin{tabular}{|c||c|c|c|} \hline\hline
  $ I_{4,f}$ & SU$(N)$ & SO$(N)$ & Sp$(N)$ \\
 \hline\hline
 $F$  & $1$  &  $1$  & $1$          \\
\hline
 $A$ & $2N$  &  $N-8$  & $N + 8$          \\
\hline
 $S_2$ & $N+8$  &  $N+8$  & $N+8$    \\
 \hline
 $A_2$ & $N-8$  & $N-8$  & $N-8$  \\
\hline\hline
\end{tabular}
\end{center}
\label{I4}
\end{table}


\begin{table}
  \caption{\footnotesize{Values of $d_R^{abcd} d_{R'}^{abcd}/d_A$
  for groups SU($N$), SO($N$), and Sp($N$) and irreducible representations $R$ 
  equal to fundamental ($F$), adjoint ($A$), and rank-2 symmetric ($S_2$) and
  antisymmetric ($A_2$) tensors.}}
\begin{center}
\begin{tabular}{|c||c|c|} \hline\hline
\multicolumn{3}{|c|}{${\rm SU}(N), \ N \geq 2$} \\
\hline
  $R$ & $d_R^{abcd} d_R^{abcd}/d_A$ & $d_R^{abcd} d_A^{abcd}/d_A$ \\
 \hline\hline
 $F$  & $\frac{N^4-6N^2 +18}{96N^2}$  &  $\frac{N(N^2+6)}{48}$   \\
\hline
 $A$ & $\frac{N^2(N^2+36)}{24}$  &  $\frac{N^2(N^2+36)}{24}$    \\
\hline
 $S_2$ & $\frac{(N+2)(N^5+14N^4+72N^3-48N^2-288N+576)}{96N^2}$  &
  $\frac{N(N+2)(N^2+6N+24)}{48}$  \\
 \hline
 $A_2$ & $\frac{(N-2)(N^5-14N^4+72N^3+48N^2-288N-576)}{96N^2}$  & 
  $\frac{N(N-2)(N^2-6N+24)}{48}$ \\
\hline\hline
\end{tabular}
\\ \vspace{1cm}
\begin{tabular}{|c||c|c|} \hline\hline
\multicolumn{3}{|c|}{${\rm SO}(N), \ N\geq 3$} \\
\hline
  $R$ & $d_R^{abcd} d_R^{abcd}/d_A$ & $d_R^{abcd} d_A^{abcd}/d_A$ \\
 \hline\hline
 $F$  & $\frac{N^2-N+4}{24}$  &  $\frac{(N-2)(N^2-7N+22)}{24}$  \\
\hline
 $A$ & $\frac{(N-2)(N^3-15N^2+138N-296)}{24}$  &
       $\frac{(N-2)(N^3-15N^2+138N-296)}{24}$   \\
\hline
 $S_2$ & $\frac{(N+2)(N^3+13N^2+110N+104)}{24}$  &
         $\frac{(N-2)(N+2)(N^2-N+28)}{24}$  \\
 \hline\hline
\end{tabular}
\qquad
\\ \vspace{1cm}
\begin{tabular}{|c||c|c|} \hline\hline
\multicolumn{3}{|c|}{${\rm Sp}(N), \ N\geq 2$} \\
\hline
  $R$ & $d_R^{abcd} d_R^{abcd}/d_A$ & $d_R^{abcd} d_A^{abcd}/d_A$  \\
 \hline\hline
 $F$  & $\frac{N^2+N+4}{384}$  &  $\frac{(N+2)(N^2+7N+22)}{384}$  \\
\hline
 $A$ & $\frac{(N+2)(N^3+15N^2+138N+296)}{384}$ & 
       $\frac{(N+2)(N^3+15N^2+138N+296)}{384}$ \\
 \hline
 $A_2$ & $\frac{(N-2)(N^3-13N^2+110N-104)}{384}$ &
         $\frac{(N+2)(N-2)(N^2+N+28)}{384}$ \\
\hline\hline
\end{tabular}
\end{center}
\label{dd}
\end{table}

\end{widetext}

\begin{thebibliography}{99}
\bibitem{fm}
%
  Since our gauge theories are vectorial, fermion mass terms are allowed by
  gauge invariance, but a fermion with nonzero mass $m$ would be integrated out
  of the low-energy effective field theory that describes the physics at
  Euclidean momentum scales $\mu < m$ and hence would not affect the infrared
  limit that we consider here. Note that we also exclude scalar fields.
\bibitem{sksb}
%
Our analysis is based on $\alpha_{IR}$ being an exact IR fixed point (IRFP)
of the renormalization group. We remark that for sufficiently smaller 
values of $N_f$, the gauge interaction produces
spontaneous chiral symmetry breaking and associated dynamical fermion 
mass generation at some scale $\Lambda$. The now-massive fermions are then
integrated out of the effective field theory that is applicable for scales 
$\mu < \Lambda$, thereby changing the beta function to that of a pure gauge
theory. Hence, $\alpha(\mu)$ evolves away from $\alpha_{IR}$, which is
thus only an approximate, but not exact, IRFP.
\bibitem{nonpos}
%
  In contrast, from our explicit calculations in \cite{dex}, we also found that
  this uniform positivity does not hold for the coefficients $d_j$ in the
  expansion of $\beta'_{IR}$ in powers of $\Delta_f$ or in the coefficients of
  the $\Delta_f$ power series for the anomalous dimension of the Dirac tensor
  fermion bilinear $\bar \psi \sigma_{\mu\nu} \psi$, where $\sigma_{\mu\nu} =
  (i/2)[\gamma_\mu,\gamma_\nu]$ is the antisymmetric Dirac tensor.
\bibitem{nfintegral}
%
Here and elsewhere, when an expression is given for $N_f$
that formally evaluates to a non-integral real value, it is understood
implicitly that one infers an appropriate integral value from it.
\end{thebibliography}
\end{document}